\documentclass[12pt]{article}

\usepackage{amsmath}         
\usepackage{amssymb}         
\usepackage{amsfonts}        
\usepackage{graphicx}        

\usepackage[table]{xcolor}
\usepackage{slashed}       
\usepackage{framed}        
\usepackage{subcaption}    
\usepackage[mathscr]{euscript} 
\usepackage{cite}          
\usepackage{booktabs}      
\usepackage{arydshln} 	    
\usepackage{tocloft}		   
\usepackage{setspace}	   
\usepackage{listings}       

\usepackage[margin=2cm]{geometry}   
\graphicspath{{figures/}}	        
\numberwithin{equation}{section}    

\newcommand{\email}[1]{\href{mailto:#1}{#1}}

\newenvironment{institutions}[1][2em]{\begin{list}{}{\setlength\leftmargin{#1}\setlength\rightmargin{#1}}\item[]}{\end{list}}

\let\oldenumerate\enumerate
\renewcommand{\enumerate}{
  \oldenumerate
  \setlength{\itemsep}{1pt}
  \setlength{\parskip}{0pt}
  \setlength{\parsep}{0pt}
}

\let\olditemize\itemize
\renewcommand{\itemize}{
  \olditemize
  \setlength{\itemsep}{1pt}
  \setlength{\parskip}{0pt}
  \setlength{\parsep}{0pt}
}

\newcommand{\MSbar}{$\overline{\text{\small MS}}$}
\newcommand{\TeV}{\ensuremath{\text{\small TeV}}}
\newcommand{\lgmus}{\ensuremath{\log\frac{\mu^2}{s}}}

\usepackage{fancyhdr}		

\usepackage[
	colorlinks=true,
	citecolor=black,
	linkcolor=black,
	urlcolor=blue,
	hypertexnames=false]{hyperref}

\fancypagestyle{firststyle}{
\rhead{\footnotesize \texttt{UCI-TR-2016-03}}}

\begin{document}

\thispagestyle{empty}
\thispagestyle{firststyle}

\begin{center}

    {\Large \bf  KK Gluons at NLO at 100 TeV}
    

    \vskip .5cm

    { \bf Benjamin~Lillard, Tim~M.~P.~Tait, Philip~Tanedo} 
    \\ \vspace{-.2em}
    { \tt
    \footnotesize
    \email{blillard@uci.edu},
    \email{ttait@uci.edu}, 
    \email{flip.tanedo@uci.edu} 
    }
	
    \vspace{-.2cm}

    \begin{institutions}[2.25cm]
    \footnotesize
    {\it 
	    Department of Physics \& Astronomy, 
	    University of California, 
	    Irvine, \textsc{ca} 92697
	    }   
    \end{institutions}
    \today
\end{center}


\begin{abstract}
\noindent 
We explore the reach of a 100 TeV proton collider to discover Kaluza-Klein gluons in a warped extra dimension. These particles are templates for color adjoint vectors that couple 
dominantly to the top quark.  We examine their production rate at NLO in the six-flavor m-ACOT scheme for a variety of reference models defining their coupling
to quarks, largely inspired by the RS model of a warped extra dimension.  In agreement with previous calculations aimed at lower energy machines,
we find that the NLO corrections are typically negative, resulting in a $K$-factor of around 0.7 (depending on the model) and with a residual scale
dependence on the order of $\pm 20\%$, greater than the variation from the scale exhibited by the na\"{i}ve LO estimate.
\end{abstract}


\renewcommand\cftsecleader{\cftdotfill{\cftdotsep}}
\renewcommand{\cftsecfont}{}

\setlength{\cftbeforesecskip}{-.2ex}




\section{Introduction}

Massive color octet vector particles (generically known as colorons, $G$)
are common ingredients in models of physics beyond the Standard Model (SM).  In particular,
in models where some or all of the SM quarks are composites, such states are ubiquitous as a consequence of the need for underlying
preon degrees of freedom which themselves carry color.  
These include topcolor models where electroweak symmetry is broken by a top condensate~\cite{Hill:1991at}, 
axigluon extensions of quantum chromodynamics with chiral symmetry breaking~\cite{Frampton:1987dn, Bagger:1987fz}, 
or technicolor models with colored composite states analogous to the $\rho$ meson~\cite{Farhi:1979zx}. 
The most popular incarnation of colorons are Kaluza-Klein (KK) excitations of the gluon in models with an extra dimension. 
As motivation, we take the particular case of the Randall--Sundrum (RS) model of a warped extra dimension~\cite{Randall:1999ee}
which is related to strong dynamics via the AdS/CFT correspondence~\cite{Maldacena:1997re, ArkaniHamed:2000ds}.
In many models, such particles have preferential coupling to the
top quark \cite{Gherghetta:2000qt,Agashe:2003zs,Agashe:2004rs,Lillie:2007hd,Kumar:2009vs}.
We explore the production of such states at a 
future 100 TeV proton--proton collider~\cite{Arkani-Hamed:2015vfh} at next-to-leading order (NLO) in quantum chromo-dynamics (QCD).
At such energies, the top quark's mass is small, leading to 
large logarithms which can be resummed into an effective top parton distribution function (PDF)
 \cite{Tung:1997xz,Brodsky:1984nx,Dawson:2014pea,Han:2014nja}.

The original RS model localized all Standard Model fields on a brane so that only gravity propagated in the bulk of the extra dimension. 
Subsequent versions of this model incorporated bulk gauge fields to alleviate constraints from proton decay and flavor-changing 
neutral currents~\cite{Chang:1999nh,Pomarol:1999ad,Davoudiasl:1999tf}, and later bulk fermions in a way that can explain the 
hierarchy in observed Yukawa couplings~\cite{Gherghetta:2000qt,Grossman:1999ra}. 
The minimal realizations of these models were tightly constrained by electroweak precision 
observables and large contributions to the $Zb\bar b$ coupling \cite{Davoudiasl:1999tf,Csaki:2002gy,Bouchart:2008vp}. 
At face value, these push the Kaluza-Klein scale to $\mathcal O(10~\text{TeV})$, 
beyond the reach of existing colliders, unless one invokes additional structure such as a gauged custodial symmetry~\cite{Agashe:2003zs,Agashe:2006at,Carena:2006bn,Carena:2007ua}
or large brane kinetic terms \cite{Davoudiasl:2002ua,Carena:2002dz,Carena:2004zn}
which allow for order TeV masses of the Kaluza-Klein (KK) excitations. 
Detailed reviews of the RS model can be found in~\cite{Csaki:2004ay, Csaki:2005vy, Ponton:2012bi, Csaki:2016kln}.

One may take the alternative viewpoint that the natural scale of RS models is $\mathcal O(10~\text{TeV})$, with a relatively modest
fine-tuning between the electroweak and compositeness scales.  From this point of view,
a more minimal model may be the realization preferred by Nature. In this case, a future 100 TeV 
collider~\cite{CEPC-SppC-PreCDR} that can access $\mathcal O(10~\text{TeV})$ 
partonic energies represents the best hope to probe the physics which underlies the electroweak scale. 
KK resonances of the gluon are likely to be the first signal of new physics as a result of their strong production cross sections. 

In these modern RS models, the Standard Model fields propagate in five-dimensional anti-de~Sitter spacetime, 
where one dimension is compact and warped. The warped dimension is an $S_1/\mathbb{Z}_2$ orbifold with fixed points, 
or ``branes," on the infrared (IR) and ultraviolet (UV) boundaries. 
The hierarchy of Yukawa couplings is suggested by the exponential profile of zero mode fermion profiles that are peaked towards 
either the infrared or ultraviolet brane according to their bulk mass parameters---corresponding to 
differing anomalous dimensions in the dual strongly coupled theory. 
The solution of the gauge hierarchy problem requires the Higgs to be largely localized on the infrared brane so that fermions which are peaked towards the infrared brane pick up large Yukawa couplings, and those peaked towards the ultraviolet brane end up with 
small Yukawa couplings. 
Further, the KK excitations of gauge bosons are redshifted by the warped background and are thus peaked towards the infrared brane. 
As such, the KK gluon has the largest wave function overlap and effective coupling to top quarks, since these are the colored 
Standard Model fermions whose wave functions are most peaked on the IR brane. The structure of the KK gluon couplings,
and its coupling to the top quark in particular, thus provides a diagnostic of RS models~\cite{Lillie:2007ve}.

In this work, we build on previous studies of KK gluons~\cite{Kong:2013xta,Agashe:2013kxa,Yu:2013wta} which studied production
at leading-order (LO) \cite{Agashe:2006hk,Lillie:2007yh,Allanach:2009vz} or NLO \cite{Chivukula:2011ng,Zhu:2012um,Chivukula:2013xla} at lower energies, where
the top content of the proton can be safely neglected.    Our aim is to provide precise estimates for the production cross section
such that detailed collider studies~\cite{Avetisyan:2013onh, Auerbach:2014xua} of the signal and background can be used to more accurately
predict the reach of a 100 TeV $pp$ machine to probe the interesting range of RS parameter space.

\section{Review of the Randall-Sundrum Framework}


The five-dimensional spacetime has a non-factorizable metric:
\begin{equation}
ds^2 = \left( \frac{R}{z} \right)^2\left( \eta_{\mu\nu} dx^\mu dx^\nu - dz^2 \right).
\end{equation}
The coordinate $x^\mu$ describes the four-dimensional Minkowski spacetime, with the metric $\eta_{\mu\nu} = \text{Diag}(+,-,-,-)$.
Coordinate $z$ describes distances in the extra dimension, and is confined to $R<z<R'$. Here, $z=R \sim 1 / M_{Pl}$ 
corresponds to the UV brane, whereas the IR brane at $z=R' \sim 1 / {\rm TeV}$ is
set by some unspecified radius stabilization mechanism.

For gauge fields, $A$, and fermions, $\Psi$, the action is given by:
\begin{equation}
S = \int\! d^4x ~dz\sqrt{g} \left[ -\frac{1}{2} F_{MN}^a F^{MN a} + i \bar\Psi \Gamma^M e^N_M D_N\Psi + i \frac{c}{R} \bar\Psi \Psi \right].
\end{equation}
The field strength tensor is defined as $F_{MN}^a = \partial_M A^a_N - \partial_N A^a_M + g_5 f^{abc} A^b_M A^c_N$, where $g_5$ is the coupling constant in the five-dimensional theory, and the vielbein is defined as $e_M^N$. 
Capital roman letters $M$ and $N$ run over all five spacetime dimensions. 
The bulk fermion mass is parameterized by a dimensionless constant $c$ times the 
AdS curvature, $k = 1/R$. 
One may also add brane-localized terms proportional to $\delta(z-R)$ or $\delta(z-R')$ to the action \cite{Davoudiasl:2002ua,Carena:2002dz}
but we neglect them for simplicity in this discussion.

In the expansion of $F_{MN} F^{MN}$ there are mixing terms between $A_\mu$ and $A_5$. To remove this mixing we choose a gauge in which $\partial_z A^\mu=0$ and $A_5 = 0$ at $z=R$ and $z=R'$, and we add gauge-fixing terms to the effective four-dimensional Lagrangian. In the general $R_\xi$ gauge, the action becomes \cite{Zhu:2011gd}:
\begin{eqnarray}
S_{5D} &=& \int\! d^4 x dz \bigg\{ \left( \frac{R}{z} \right)^4 \bar \Psi \left[ g_5 \gamma^\mu A_\mu + i g_5 \gamma^5 A_5 + i \frac{c}{R} \right] \Psi  \label{eq:FiveDAction}
\\&&\ + \frac{R}{z} \left(-\frac{1}{2\xi} \right) \left[ \partial^\mu A^a_\mu - \xi \left( \frac{z}{R} \right) \partial_z \left( \frac{R}{z} A_5 ^a \right) \right]^2 + \frac{R}{z} \bar c^a \left[ -\partial^\mu D_\mu + \xi(\frac{z}{R}) \partial_z \frac{R}{z} \partial_z \right]^{ab} c^b \bigg\}. \nonumber
\end{eqnarray}
We work in the Feynman gauge, $\xi=1$.

\subsection{Kaluza-Klein decomposition}

\paragraph{Vector bosons}
A five-dimensional bulk gauge field can be decomposed into orthogonal functions. 
\begin{eqnarray}
A_\mu^a(x,z) &=& \frac{1}{\sqrt{R}} \sum_{j=0} A_\mu^{a(j)}(x) f_j (z) \\
A_5^a (x,z) &=& \frac{1}{\sqrt{R}} \sum_{j=1} A_5^{(j)}(x) \frac{1}{m_j} \partial_z f_j(z).
\end{eqnarray}
The wave functions $f_j(z)$ can be expanded in terms of Bessel functions, and satisfy:
\begin{align}
\int\!\frac{dz}{z} f_i (z) f_j (z) = \delta_{ij}
\;;&& \;
\partial_z \left( \frac{1}{z} \partial_z f_j \right) + \frac{m_j^2}{z} f_j = 0
\;; &&\;
\left. \partial_z f_j(z) \right|_{z=R,R'} = 0 .
\end{align}
Theories with unbroken gauge bosons have flat zero modes, with $\partial_z f_0(z) = 0$,
\begin{equation}
f^{0}(z) \equiv f_0 = \sqrt{\frac{1}{\log(R'/R)}}.
\end{equation}
For SU(3)$_c$ this zero-mode is identified with the Standard Model (QCD) gluon, and we refer to higher modes as KK gluons.

\paragraph{Chiral fermions}
In four dimensional spacetime, the Standard Model fermions are left- or right-chiral Weyl fermions. In five dimensions, the 
$\gamma^5$ matrix is appropriated into $\gamma^M = (\gamma^0, \ldots, \gamma^3, \gamma^5)$ and the bulk quarks and leptons are 
four-component Dirac spinors. To recover the chiral Standard Model, 
we impose boundary conditions to remove the wrong-chirality states for the fermion zero-modes.

The Weyl components of the Dirac spinor $
\Psi = \left( \begin{array}{c} \chi \\ \bar\psi \end{array} \right)
$
are expanded separately as:
\begin{align}
\chi(x,z) =\frac{1}{R} \sum_{j=0} h_L^j(z) \chi_j(x) 	&, & \bar\psi (x,z) = \frac{1}{R} \sum_{j=0} h_R^j(z) \bar \psi_j(x),
\end{align}
with the orthogonality relations:
\begin{align}
\int\! \frac{dz}{R} \left(\frac{R}{z} \right)^4 h_L^i (z) h_L^j(z) \;=\; \int\! \frac{dz}{R} \left(\frac{R}{z} \right)^4 h_R^i (z) h_R^j(z) \;=\; \delta_{ij}.
\end{align}
For $c<-1/2$ the zero mode peaks towards the UV boundary; for $c>-1/2$, it peaks towards the IR.
An anarchic flavor model with $\mathcal O(1)$ Yukawa couplings in the 5D Lagrangian suggests~\cite{Lillie:2007hd}
$c_{t_R} \approx 0$, $c_{Q_{3L}} \approx 0.4$, and $c_f < -0.5$ for all other quarks to reproduce the observed hierarchy in their masses. 
In this way the RH top quark peaks strongly to the IR brane, and the LH $Q_3$ doublet is relatively flat.

\subsection{Interactions in four dimensions}
Interactions between particular KK modes of the 
fermions and bosons can be derived from the five-dimensional theory by integrating over $z$.
From the 5D action, Eq.~(\ref{eq:FiveDAction}) we determine the relevant Feynman rules for the effective 4D theory
describing the $(0)$ and $(1)$ KK modes.  Because the gauge boson zero modes are flat and the functions $f^{(n)}$ are orthogonal, some couplings vanish.

\paragraph{QCD gluon couplings}
To relate $g_5$ of the 5D theory to the 4D coupling $g_s$, we extract the three-gluon vertex for the gauge boson zero mode.
\begin{eqnarray} 
\mathcal L_{3g} &=& \int\! dz \frac{R}{z} \left( -\frac{1}{2} g_5 f^{abc} A_\mu^b A_\nu^c (\partial_\mu A_\nu^a - \partial_\nu A_\mu^a ) \right) \left[\frac{ f_0 }{\sqrt{R}}\right]^3 \\
g_s &=& \frac{f_0 g_5}{\sqrt{R} }.
\end{eqnarray}
With the definition of $g_s$ above, the Feynman rules for the three-point vertex with zero-mode gluons matches QCD.
Using the orthogonality of the $f_i$ and $h_{L,R}^i$ basis functions, it can be shown that the zero-mode gluon couples to fermions and other KK gluon modes with the same coupling $g_s$,
\begin{align}
g_5 \int\!dz \frac{R}{z} \left[ \frac{1}{\sqrt{R}^3} f_0(z) f_i(z) f_j(z) \right] \;=\; \frac{1}{\sqrt{R}} f_0 g_5 \int\frac{dz}{z} f_i(z) f_j(z) 
\;=\; \frac{f_0}{\sqrt{R}}  g_5 \delta_{ij} &=\; g_s \delta_{ij}; \\
g_5 \int\!dz \left(\frac{R}{z} \right)^4 \left[ \frac{1}{\sqrt{R}} f_0(z) h_{L,R}^i(z) h_{L,R}^j(z) \right] \;=\; g_5 \frac{f_0}{\sqrt{R}} \int\!dz \left(\frac{R}{z} \right)^4 h_{L,R}^i(z) h_{L,R}^j(z) 
&=\; g_s \delta_{ij},
\end{align}
as demanded by gauge invariance.

\paragraph{KK gluon couplings}
We are primarily interested in the coupling of the KK gluon to the left- and right-handed fermions. These stem from the terms in the action:
\begin{equation}
\int\! d^4 x \left[ g_L \bar\chi^{(0)} \bar\sigma^\mu A_\mu^{(1)} \chi^{(0)} + g_R \psi^{(0)} \sigma^\mu A_\mu^{(1)} \bar\psi^{(0)} \right],
\end{equation}
leading to couplings,
\begin{eqnarray}
g_L  &=& \frac{g_5}{\sqrt{R}}  \int\! dz \left(\frac{R}{z} \right)^4 (h_L^0)^2(z) f_1(z), \\
g_R &=& \frac{g_5}{\sqrt{R}}  \int\! dz \left(\frac{R}{z} \right)^4 (h_R^0)^2(z) f_1(z).
\end{eqnarray}

\subsection{Feynman rules}
\begin{figure}[t]
\centering
\includegraphics[scale=1.0]{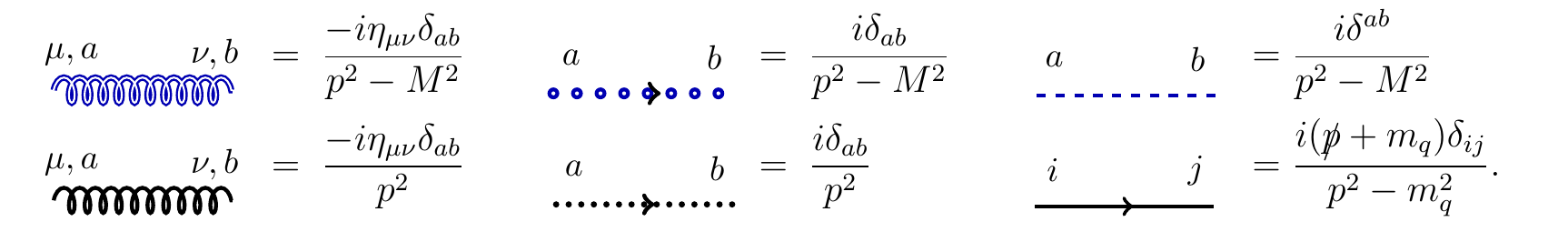}
\caption{Propagators, Feynman gauge. Top: KK gluon, KK ghost, and $A_5^{(1)}$. Bottom: QCD gluon, QCD ghost, and SM quark.} \label{figure:rules:propagators}
\end{figure}

In this subsection, we summarize the Feynman rules needed for our calculation. Figure~\ref{figure:rules:propagators} 
shows the propagators in the $\xi=1$ gauge. Typically the quark mass $m_q$ will be set to zero 
(except in Section~\ref{section:massivequark}). The KK gluon mass is denoted by $M$.

The leading-order process is determined by the interaction of SM quarks with the KK gluon:
\begin{align}
\raisebox{-.48\height}{\includegraphics{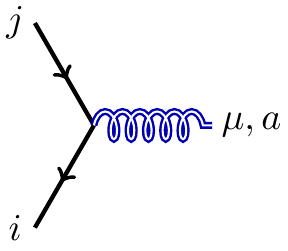}}
\;
&=\; i \gamma^\mu  T^a (g_L P_L + g_R P_R)
\;
= \; i \gamma^\mu(g_V + g_A \gamma_5) T^a .
\end{align}
where $T^a$ indicates a generator of SU(3) in the fundamental representation.
The virtual $\mathcal O(\alpha_s)$ corrections also involve the triple gluon $(0)$-$(1)$-$(1)$ vertex, and
interactions with the $A_5^{(1)}$ scalar and Fadeev-Popov ghosts:
\begin{align*}
\raisebox{-.48\height}{\includegraphics{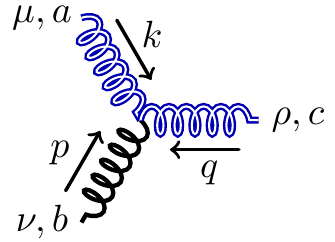}}\;
&= g_s f^{abc} \left[ \eta^{\mu\nu} (k-p)^\rho + \eta^{\nu\rho} (p-q)^\mu + \eta^{\rho \mu} (q- k)^\nu \right].
\end{align*}
\begin{align*}
\raisebox{-.48\height}{\includegraphics{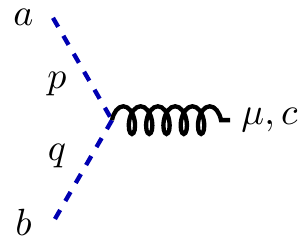}}
\;
&= -g_s f^{abc} (p - q)^\mu 
\; &&\;
\raisebox{-.48\height}{\includegraphics{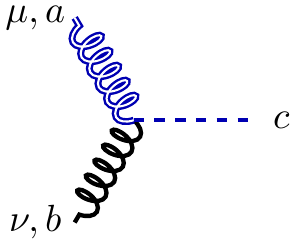}}
\;
= -i M g_s f^{abc} \eta^{\mu\nu} 
 \\ \\
\raisebox{-.48\height}{\includegraphics{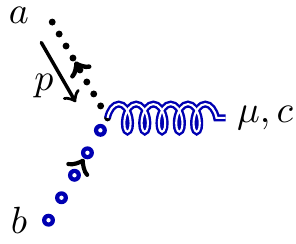}}
\;
&= -g_s f^{abc} p^\mu 
\; &&\;
\raisebox{-.48\height}{\includegraphics{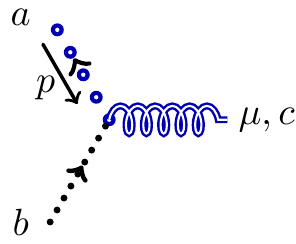}}
\;
= -g_s f^{abc} p^\mu 
\end{align*}
where all momenta flow into the vertex and $f^{abc}$ are the structure constants.  The 
pure QCD interactions involving only zero modes are unchanged with respect to the Standard Model. 

\paragraph{Other interactions}
It can be shown that the $A_5$ couples to quarks with an interaction proportional to the quark mass. 
In Section~\ref{section:higherorder} we justify neglecting these corrections.
\begin{align}
\raisebox{-.45\height}{\includegraphics{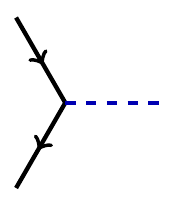}}
\;
= \frac{m_q}{M}  T^a (g_L - g_R )\gamma_5
\; \rightarrow 0.
\end{align}
The three-point interaction involving three KK gluons is proportional to:
\begin{align}
g_{(111)} &\equiv \frac{g_5}{\sqrt{R}} \int\! \frac{dz}{z} \left(  f^{(1)}(z) f^{(1)}(z) f^{(1)}(z) \right) \nonumber
\\
\raisebox{-.45\height}{\includegraphics{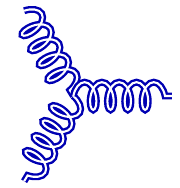}}
\;
&= g_{(111)} f^{abc} \left[ \eta^{\mu\nu} (k-p)^\rho + \eta^{\nu\rho} (p-q)^\mu + \eta^{\rho \mu} (q- k)^\nu \right].
\end{align}
This coupling is not necessarily small, but we do not include it in the present calculation. It was calculated in~\cite{Allanach:2009vz} for typical values of $R$ and $R'$ with the result $g_{(111)} \simeq 2.5 g_s$.

\section{KK Gluon Production}

At a hadron collider, the leading order process for KK gluon production is through annihilation of two initial state quarks
of the same flavor, 
$q\bar q\rightarrow G$.  The fact that the heavy quarks 
typically have larger couplings than the lighter quarks is balanced by the fact that 
heavy quarks in the proton are derived from collinear gluon splitting. 
At energies much higher than the quark mass, gluon splitting is enhanced by large logarithms which are resummed into a heavy quark PDF.  
For a KK gluon with $M\simeq \mathcal O(10~\TeV)$, even the top quark can be treated as effectively massless for most purposes.

\begin{figure}
\centering
\includegraphics[scale=1.0]{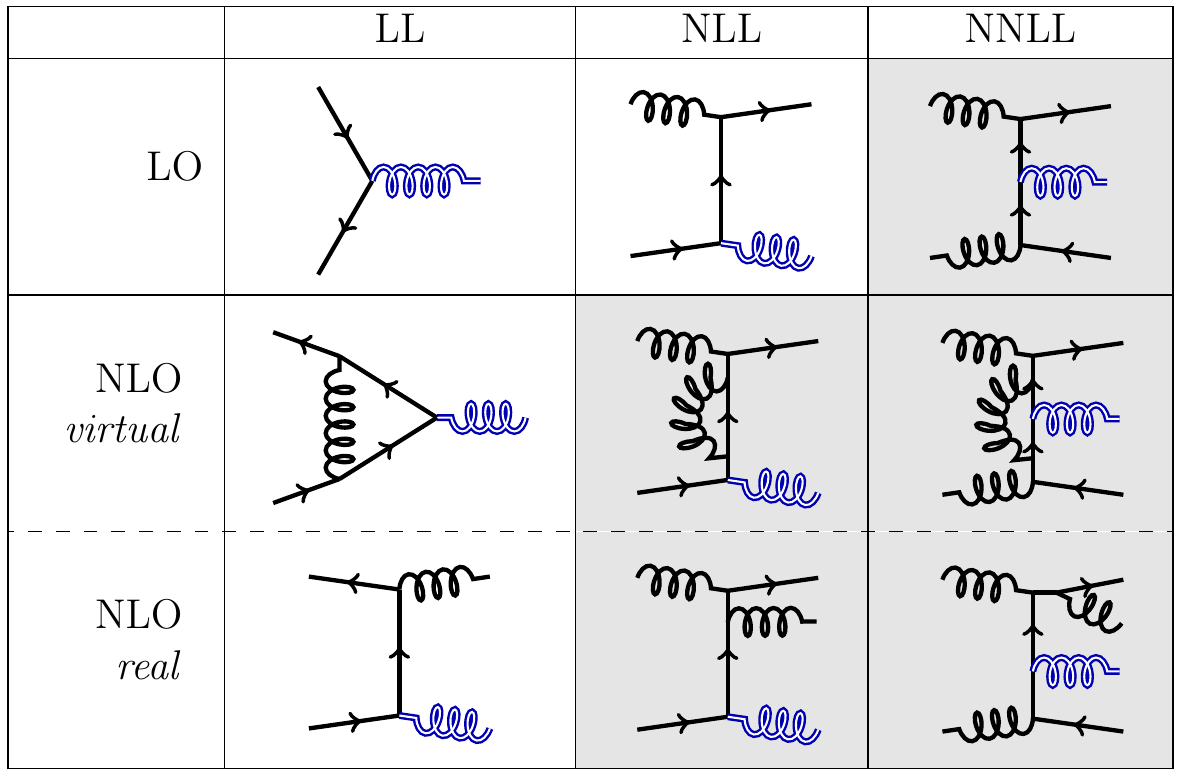}
\caption{Perturbative expansion in $\alpha_s$ and $\log(\mu^2/m_t^2)$. We include the NLO/LL and LO/NLL diagrams in our calculation, and we neglect the shaded corrections above.}
\label{figure:loNLOnll}
\end{figure}

Since $\alpha_s \sim 0.1$ at TeV energies, higher order corrections to KK gluon production
are generically important.  In Figure~\ref{figure:loNLOnll}, we show 
representative diagrams for the LO and NLO contributions to KK gluon production
(as well as higher order diagrams in the shaded portion).  At NLO
there are both virtual corrections to the LO process, as well as real corrections where an additional parton is radiated into the
final state.  When the radiated particle is a light quark or gluon, the correction is $ \mathcal O(\alpha_s)$ compared to the
leading order diagram.  When it is a heavy quark, the LO term already implicitly contains a factor of $\alpha_s \log (\mu^2 / m_t^2)$ from each gluon splitting kernel. Compared to the LO, real corrections with a $(q_i g)$ initial state are effectively of order $\log^{-1} (\mu^2 / m_t^2)$, 
after subtracting the appropriate counter-term in the ACOT~\cite{Aivazis:1993pi} scheme to avoid double-counting the collinear regime.
We refer to processes with one initial gluon as ``next-to-leading log" (NLL) and those with two initial gluons as ``next-to-next-to-leading log" (NNLL).

\subsection{Leading Order Cross Section}

In general, the KK gluon couples chirally to the quarks.  In a typical RS model,
this is primarily important for the top and bottom quarks.  The matrix element is given by:
\begin{align}
\raisebox{-.48\height}{\includegraphics{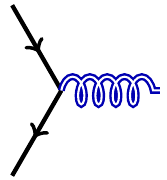}}
\; = \; i \mathcal M_L + i \mathcal M_R 
\;=\; i \bar v_2 \gamma^\mu (g_L P_L + g_R P_R) T^a u_1 \epsilon^\star_{3\mu}
\;=\; i \mathcal M_\text{LO}.
\end{align}
It is useful to define:
\begin{align}
\frac{g_L+g_R}{2} \equiv g_V
\;
\ \ \ \ \ 
\;
\frac{g_R - g_L}{2} \equiv g_A.
\end{align}
Note that for the light quarks, typically $g_A\approx 0$. 
In the cross section, the average over initial states includes 2 spins and $N_c$ colors for each quark.
It will be convenient to leave 
$N_c=3$ and $C_F = 4/3$ explicit in the calculation.  The integration over phase space is trivial, leading
leading to a LO partonic cross section of
\begin{eqnarray}
\sigma_\text{LO} &=& \frac{\pi}{s} \left| \mathcal M_\text{LO} \right|^2 \delta(s-M^2) 
~=~ \sigma_0 (1-\epsilon)~ \frac{\delta(1-\tau)}{M^2},
\label{eq:sigmaLO}
\end{eqnarray}
where we define $\tau ~\equiv~ M^2/s$, and:
\begin{equation}
\sigma_0 \; \equiv \; (g_V^2 + g_A^2) \frac{\pi N_c C_F}{N_c^2} \; =\;  \frac{\pi C_F }{2 N_c} (g_L^2 + g_R^2). 
\end{equation}

\subsection{Higher Order Corrections}	\label{section:higherorder}
In Section~\ref{section:virtual} we compute virtual corrections to the KK gluon production to order $\alpha_s$, $g_L^2$ and $g_R^2$.  We neglect the order
$g_{(111)}^2$ corrections, which are typically subdominant and not enhanced by large logarithms~\cite{Chivukula:2013xla}.
These contributions typically contain ultraviolet (UV) and infrared (IR) divergences, which we regulate with dimensional regularization. 
We renormalize in the \MSbar\ subtraction scheme to remove the UV divergences.

In Section~\ref{section:real} we calculate the real corrections from $2\rightarrow2$ processes such as $q_i\bar q_i \rightarrow G g$ and $q_i g \rightarrow q_i G$. 
These include IR and collinear divergences, which cancel between the virtual corrections, the real emission contributions, and the PDF counter-terms. 
The divergences cancel independently for each distinct initial state ($q_i \bar q_i $, $g q_i$, $g \bar q_i$), allowing us to consider the NLO and NLL perturbations separately.

In the virtual corrections and some of the real corrections we omit the mass of the top quark, as in the simplified ACOT 
scheme (s-ACOT) \cite{Kramer:2000hn}.
This is not necessarily appropriate for the NLL cross section $t g \rightarrow t G$, which includes diverging logarithms in the $s\rightarrow M^2$ limit. 
We follow the modified ACOT scheme (m-ACOT) of~\cite{Han:2014nja}, in which the top quark mass is retained in the $t g \rightarrow t G$ 
cross section to regulate the collinear divergence.
We show in Section~\ref{section:massivequark} that although s-ACOT and m-ACOT lead to different expressions for the NLL 
cross section, the effect on the total cross section is not large.

\section{Virtual Corrections}	
\label{section:virtual}

At this order, the virtual corrections take the form of a one-loop diagram interfering with the leading order graph, and share its
$2 \rightarrow 1$ kinematics.  In the process at hand, they can be divided into self-energy corrections and corrections to the vertex.

The relevant part of the renormalized Lagrangian describing the KK gluon and fermion zero modes can be written,
\begin{eqnarray}
\mathcal L_{f+\text{KK}} &=&- \frac{1}{4} (\partial_\mu A_\nu^a - \partial_\nu A_\mu^a)^2 -\frac{ M^2}{2} A_\mu^a A^{a\mu} -\frac{\delta M^2}{2} A_\mu^a A^{a\mu} - \delta_1 \frac{1}{4} (\partial_\mu A_\nu^a - \partial_\nu A_\mu^a)^2  
 + \sum_{j=L,R}   \bigg[ \bar \Psi_{j}i \slashed \partial \Psi_{j} \nonumber\\&&\  
 +  \bar \Psi_{j} \gamma^\mu \left[g_L P_L + g_R P_R \right] T^a \Psi_{j} A_\mu^a 
 + \delta_Q^{(j)} \bar \Psi_{j} i \slashed \partial \Psi_{j}  
  + \delta_{j} \bar \Psi_{j} \gamma^\mu \left[g_L P_L + g_R P_R\right] T^a \Psi_{j} A_\mu^a 
\bigg], 
\end{eqnarray}
where the counter-terms are related to the wave function renormalization constants in the usual way,
\begin{align}
\delta_Q^{L,R} = Z_Q^{L,R} - 1 &&
\delta_1 = Z_1 -1 && 
\delta M^2 = Z_1 M_0^2 - M^2
\end{align}
\begin{align}
\delta_{L,R} = Z_{L,R} Z_Q^{L,R} \sqrt{Z_1} - 1 = 
\delta Z_3^{L,R} + \delta_Q^{L,R} + \frac{1}{2} \delta_1 .
\end{align}
As shown in detail below in Section~\ref{sec:selfenergy}, the counter-terms are determined in terms of the one loop self-energy
diagrams in the \MSbar\ scheme.

In the \MSbar\ scheme, the propagators do not generically have poles with unit residue on-shell.  As a result,
there is a contribution from the self-energy diagrams through the LSZ reduction.  We denote the amount by which
the residues differ from one by $\delta R_Q^{L,R}$ and $\delta R_1$ (computed below), respectively.  At NLO, the amplitude
for $q\bar q \rightarrow G$ can be written:
\begin{eqnarray}
i\mathcal M &=& i \sqrt{R_1} \Big( R_Q^L \mathcal M_L + R_Q^R \mathcal M_R + \mathcal M_{NLO}^{\text{vertex}} + ... \Big)\\
& \simeq & i \Big(1 + \delta R^L_Q+ \frac{1}{2} \delta R_1\Big) \times \mathcal M_L 
~+~ i \Big(1 + \delta R^R_Q+ \frac{1}{2} \delta R_1\Big) \times \mathcal M_R
~+~ i \mathcal M_{NLO}^{\text{vertex}}. 		
\label{eq:amplitude:LSZ}
\end{eqnarray}
After renormalizing the couplings this expression will be UV-finite, but will still contain residual soft divergences 
that will cancel those from the $2\rightarrow2$ gluon emission process.

\subsection{Self-energy Corrections}
\label{sec:selfenergy}

In this section, we compute the self-energy corrections to the quarks and to the KK gluon in order to extract the 
order $\alpha_s$ corrections to the residues
$\delta R^{L,R}_Q$ and $\delta R_1$ in the \MSbar\ scheme. 
 After renormalization, these will be UV finite (but generically still IR-divergent).

\subsubsection{Quark Self-energy}

The quark self-energy receives corrections at $\mathcal O(\alpha_s)$ from the zero-mode gluon, 
and others proportional to $g_L^2$ and $g_R^2$ from the KK gluon. 
The counter-terms $\delta^{L,R}_Q$ cancel the UV divergences of the fermion wave function.
\begin{align}
\raisebox{-.48\height}{\includegraphics{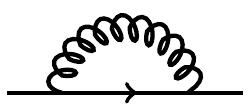}}
\;
& \equiv \; -i \Sigma_A(\slashed p) \; = \; ig_s \gamma^\mu T^a_{ij} \int\! \frac{d^d k}{(2\pi)^d} \frac{-i}{k^2} \frac{i (\slashed k + \slashed p)}{(k+p)^2} i g_s \gamma_\mu T^a_{jk} \\[0.5cm] 
\raisebox{-.48\height}{\includegraphics{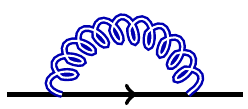}}
\;
& \equiv \; -i \Sigma_B(\slashed p) \; = \; \int\!\frac{d^d k}{(2\pi)^d} i \gamma^\mu (g_L P_L + g_R P_R) T^a \frac{i(\slashed k + \slashed p)}{(k+p)^2} \frac{-i \cdot i \gamma_\mu}{k^2-M^2} (g_L P_L + g_R P_R) T^a
\end{align}
This leads to the following UV divergences:
\begin{align}
 -i\Sigma_A^\text{UV} & = \; \frac{i \slashed p \alpha_s}{4\pi} C_F \Big( \frac{1}{\bar\epsilon} \Big). \\
 -i\Sigma_B^\text{UV} & = \; \frac{ i \slashed p C_F }{(4\pi)^2} \frac{1}{\bar\epsilon} (g_L^2 P_L + g_R^2 P_R). 
\end{align}
The \MSbar\  counter-terms are thus:
\begin{align}
\delta_Q^{L} \;=\; -\frac{C_F}{\bar\epsilon} \left[ \frac{\alpha_s}{4\pi} + \frac{g_L^2 }{(4\pi)^2} \right],
&&
\delta_Q^R \; =\; -\frac{C_F}{\bar\epsilon} \left[ \frac{\alpha_s}{4\pi} + \frac{g_R^2}{(4\pi)^2} \right].
\end{align}
The corrections to the residues $\delta R^{L,R}_Q$ are extracted from the derivative of the self-energy:
\begin{eqnarray}
-i\Sigma_A(\slashed p) 
&=& g_s^2 (2-2\epsilon) C_F \int_0^1\! dx (1-x) \int\!\frac{d^d \ell}{(2\pi)^2} \frac{\slashed p}{(\ell^2 -\Delta_A)^2}, \; \; \; \; 
 \Delta_A =  p^2 (x^2-x)\\
-i \frac{d}{d\slashed p} \Sigma_A(\slashed p) &=& i g_s^2 (2-2\epsilon) C_F \int_0^1\! dx (1-x) \int\!\frac{d^d \ell}{(2\pi)^2} \Big[ \frac{1}{(\ell^2-\Delta)^2} - \frac{4p^2 x(1-x)}{(\ell^2 - \Delta)^3} \Big] \\
\frac{d\Sigma_A(\slashed p=0)}{d\slashed p} &=& 0.
\end{eqnarray}
In the on-shell limit, the loop integrals become scaleless: $\Delta_A = -p^2 x(1-x) \rightarrow 0$. 
As shown in Appendix~\ref{section:scalelessloop}, the IR and UV divergences precisely cancel each other.
This is not the case with the KK gluon loop, which is not IR divergent:
\begin{align}
-i \frac{d \Sigma_B}{d \slashed p} &= \frac{i C_F}{(4\pi)^2} \Gamma(\epsilon) (2-2\epsilon)  \int_0^1\! dx (1-x)^{1-\epsilon} \left( \frac{4\pi \mu^2}{M^2 - x^2 p^2} \right)^\epsilon \left[ 1+ \frac{2\epsilon x p^2 }{M^2 - xp^2} \right] (P_L g_L^2 + P_R g_R^2) \\
\frac{d \Sigma_B( \slashed p =0)}{d \slashed p} &= - C_F \frac{P_L g_L^2 + P_R g_R^2}{(4\pi)^2} \left[ \frac{1}{\bar\epsilon} + \log\frac{\mu^2}{M^2} - \frac{1}{2} \right].
\end{align}
Finally, we add the contribution from the counter-terms,
\begin{eqnarray}
\frac{d}{d\slashed p} \Sigma_\text{CT} 
&=& 
\; - \delta^{L}_Q P_L -  \delta^{R}_Q P_R. \\
\frac{d}{d\slashed p }\left[ \Sigma_A + \Sigma_B + \Sigma_\text{CT} \right] &=& C_F \frac{P_L g_L^2 + P_R g_R^2}{(4\pi)^2} \left[- \frac{1}{\bar\epsilon} - \log\frac{\mu^2}{M^2} + \frac{1}{2} + \frac{1}{\bar\epsilon} \right] 
+ \frac{C_F}{\bar\epsilon} \left[ \frac{\alpha_s}{4\pi} \right].
\end{eqnarray}

\paragraph{Fermion residue:}
The residue of the full propagator of the renormalized fermion field is $R_Q^{L,R} \equiv 1+\delta R_Q^{L,R}$, with:
\begin{eqnarray}
\delta R_Q^{L,R} &=& \frac{\alpha_s}{4\pi} \frac{C_F}{\bar\epsilon} + \frac{ g_{L,R}^2 }{(4\pi)^2} C_F \left[ \frac{1}{2} - \log\frac{\mu^2}{M^2}  \right].
\label{eq:residue:fermion}
\end{eqnarray}
Note that the correction to the residue still includes an IR divergence.

\subsubsection{KK Gluon Self-energy}

The $\mathcal O(\alpha_s)$ corrections include gluons, ghosts, and the $A_5$ scalar.
Corrections from quark loops are $\mathcal O(g_L^2 + g_R^2)$ rather than $\mathcal O(\alpha_s)$. 
\begin{align}		\label{eq:selfenergy:KK}
\raisebox{-.0\height}{\includegraphics[scale=0.96]{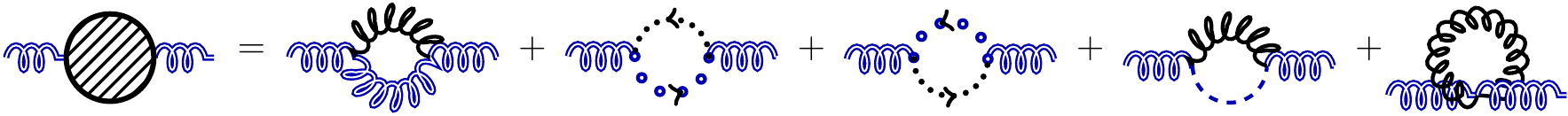}}
 .
\end{align}
The contribution from the counter-terms takes the form:
\begin{align}
\raisebox{-.44\height}{\includegraphics{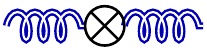}}
\;
= i \delta^{ab} \Big( \eta^{\mu\nu} (\delta_1 q^2 + \delta M^2 ) -\delta_1 q_\mu q_\nu \Big). 	\label{eq:counterterm:gluonfield}
\end{align}
The functions $\Pi(q^2)$ and $\Delta(q^2)$ denote the coefficients of the two tensor forms that appear in the two-point function:
\begin{align}
\raisebox{-.44\height}{\includegraphics{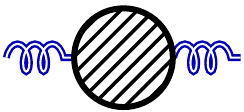}}
\;
\equiv
\;
i \Pi^{\mu\nu}=i\Big(\Pi(q^2) \eta^{\mu\nu} - \Delta(q^2) q^\mu q^\nu\Big) .		
\end{align}
%

\paragraph{Resumming the two-point functions:}
Because of the longitudinal polarizations, resumming the two-point corrections is slightly more complicated. 
In the Feynman gauge the propagator has a simple form, allowing us to write:
\begin{equation}
i\Pi^{\mu\alpha} \frac{-i \eta_{\alpha\beta}}{q^2 - M^2 + i\varepsilon} = \frac{\Pi \eta^\mu_\beta - \Delta q^\mu q_\beta}{q^2 - M^2 + i\varepsilon} \equiv \mathcal T^\mu_\beta.
\end{equation}
This is the tensor that appears for every additional two-point function added to the propagator. ``Squaring" this tensor produces:
\begin{eqnarray}
\mathcal T^\mu_\alpha \mathcal T^\alpha_\beta \;=\; 
\big(\mathcal T^2\big)^\mu_\beta &=& \Big(\frac{1}{q^2 - M^2} \Big)^2 \Big[\Pi^2 \eta^\mu_\beta -\Big(2\Pi\Delta - q^2\Delta^2\Big) q^\mu q_\beta \Big].
\end{eqnarray}
For massless (and therefore transverse) bosons, $\Pi = q^2 \Delta$, and the $q^\mu q_\beta$ term simplifies so that everything is proportional to $\Pi^2$. For a massive boson this is not generally true.  However,
this potentially messy remainder only shows up in the $q^\mu q_\beta$ term 
and disappears when contracted with the on-shell fermion bilinear $\bar v_2 \gamma^\mu u_1$:
\begin{equation}
f(q^2) q^\mu \bar v_2 \gamma^\mu u_1 = f(q^2) \bar v_2 \slashed q u_1 = f(q^2) \bar v_2 (m_f - m_f) u_1 = 0.
\end{equation}
Thus, for our purpose it is sufficient to consider only the $\Pi(q^2) \eta^{\mu\nu}$ part of the two-point function.
In this case, the full propagator becomes:
\begin{eqnarray}
\frac{-i \eta_{\mu\nu} }{q^2 - M^2} + \frac{-i\eta_{\mu\alpha}}{q^2-M^2} \frac{ \mathcal T^\alpha_\nu }{q^2 - M^2} + \ldots &=& \frac{-i \eta_{\mu\nu}}{q^2-M^2 - \Pi(q^2)} + f(q^2)q_\mu q_\nu.
\end{eqnarray}
From here on, we drop the $q_\mu q_\nu$ term.

\paragraph{Two-point function:}
We label the various contributions to $\Pi(q^2)$ as $A$, $B$, $C$, and so on by their order of appearance in~(\ref{eq:selfenergy:KK}). Based on the reasoning of the previous section, we can discard any terms proportional to $q^\mu$ or $q^\nu$.
%
%
\begin{eqnarray}
i\Pi_A \eta^{\mu\nu} \delta^{ab} &=& \int\! \frac{d^d k}{(2\pi)^d} g_s f^{acd} \Big( \eta^{\mu\alpha} (q-k)^\beta + \eta^{\alpha\beta} (2k + q)^\mu + \eta^{\beta\mu} (-k-2q)^\alpha \Big) \frac{-i\eta_{\alpha A}}{k^2}\nonumber\\&&\ \times
  \frac{-i \eta_{\beta B}}{(k+q)^2 - M^2} g_s f^{bdc} \Big( \eta^{\nu A} (q-k)^B + \eta^{A B} (2k + q)^\nu + \eta^{B\nu} (-k-2q)^A \Big) \nonumber \\
&=& +g_s^2 N_c \delta^{ab} \int\! \frac{d^d k}{(2\pi)^d} \frac{\eta^{\mu\nu} (2k^2 + 2k\cdot q + 5q^2 ) + 2 k^\mu k^\nu (5 - 4\epsilon) }{k^2 [(k+q)^2-M^2]}.
\end{eqnarray}
%
%
The two ghost diagrams contribute equivalent terms to $\Pi$:
\begin{eqnarray}
i\Pi_B\eta^{\mu\nu} \delta^{ab} =  i\Pi_C\eta^{\mu\nu} \delta^{ab}
&=& (-1) \int\! \frac{d^d k}{(2\pi)^d} \frac{g_s f^{adc} (q+k)^\mu g_s f^{bcd} k^\nu \cdot i \cdot i} {k^2 [(q+k)^2 - M^2] } \\
&=&-g_s^2 N_c \delta^{ab} \int\! \frac{d^d k}{(2\pi)^d} \frac{k^\mu k^\nu}{k^2 [(q+k)^2 - M^2]}.
\end{eqnarray} 
%
%
From the $A_5$ diagram:
\begin{eqnarray}
i\Pi_D \eta^{\mu\nu} \delta^{ab} &=&\int\! \frac{d^d k}{(2\pi)^d} \frac{(-ig_s M f^{adc} \eta^{\mu\alpha} )(-i \eta_{\alpha\beta})\cdot i\cdot (-ig_s M f^{cbd} \eta^{\beta\nu} ) }{k^2 [(q+k)^2 - M^2]} \\
i\Pi_D &=& -g_s^2 M^2 N_c \int\! \frac{d^d k}{(2\pi)^d} \frac{1}{k^2 [(q+k)^2 - M^2]}.
\end{eqnarray}
%
%
Diagram E is the easiest to calculate, being scaleless and thus vanishing in dimensional regularization:
\begin{equation}
i\Pi_E = 0.
\end{equation}
%
%
We simplify $k^\mu k^\nu$ using the symmetric loop momentum $\ell\equiv k - qx$, 
discarding $q^\mu$ and $q^\nu$, and replacing $\ell^\mu \ell^\nu$ with $\eta^{\mu\nu}$.
%
After this replacement, the total $\mathcal O(\alpha_s)$ two-point function for the KK gluon is: 
\begin{eqnarray}
%
%
%
%
\Pi(q^2)&=& \frac{N_c g_s^2 \Gamma(\epsilon)}{(4\pi)^2} \int^1_0\! dx \Big(\frac{4\pi \mu^2}{\Delta} \Big)^\epsilon \Big[\frac{8-6\epsilon}{1-\epsilon} \Delta + \Big( 5 - 2x + 2x^2 \Big) q^2 - M^2 \Big],	\label{eq:Pi:int}
\end{eqnarray}
where in this expression:
\begin{equation}
\Delta \equiv x(M^2 - q^2) + x^2 q^2.
\end{equation}
%
%
%
%
Quark loop corrections add the following diagram to the KK gluon self-energy:
\begin{align*}
\raisebox{-.0\height}{\includegraphics{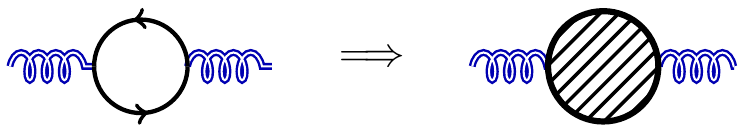}}
\end{align*}
This adds a UV-diverging, IR-finite piece to the two-point function.
By simplifying the projector matrices $P_L$ and $P_R$, this self-energy amplitude is: 
\begin{eqnarray}
i \Pi_F^{\mu\nu} \delta^{ab} 
&=& (-1) \frac{\delta^{ab} }{2} \left( \frac{g_L^2 + g_R^2 }{2} \right) \int_0^1 \! dx \int\! \frac{d^d k}{(2\pi)^d} \frac{\text{Tr}[ \gamma^\mu \slashed k \gamma^\nu (\slashed k + \slashed q ) ] }{D^2} ,
\end{eqnarray}
where $D = k^2 + 2x k\cdot q + x q^2 \, \equiv \, \ell^2 - \Delta$, with $\Delta = - x(1-x) q^2$. After discarding the $q^\mu q^\nu$ terms, we can write:
\begin{eqnarray}
i \Pi_F^{\mu\nu} &=& \eta^{\mu\nu} \sum_{i=1}^{6} (g_L^2 + g_R^2) \frac{ i \Gamma(\epsilon) }{(4\pi)^2} \int_0^1 \! dx \left( \frac{4\pi \mu^2 }{\Delta} \right)^\epsilon (-2 q^2) (x - x^2).
\end{eqnarray}
This sum over quark couplings can be expressed as a constant, defined as:
\begin{equation}
G_T^2 \equiv \sum_{i = 1}^6 \left[ (g_L^{(i)} )^2 + (g_R^{(i)} )^2 \right].
\end{equation}

\paragraph{Renormalization:}
All together, the UV divergent part of the KK gluon self-energy is:
\begin{eqnarray}
\Pi_{UV} &=& \frac{N_c \alpha_s}{4\pi} \frac{1}{\bar\epsilon} \int^1_0\! dx \Big[8 x(M^2 - q^2) + 8 x^2 q^2 + \Big(5-2x+2x^2\Big)q^2 - M^2 \Big] \nonumber\\&&\
+ \frac{G_T^2}{(4\pi)^2} (-2 q^2) \frac{1}{\bar\epsilon}  \int_0^1 \! dx (x - x^2) \\
\Pi_{UV} &=& \frac{N_c \alpha_s}{4\pi} \frac{1}{\bar\epsilon} \Big[ 3M^2 + \frac{10}{3} q^2 \Big] 
+  \frac{G_T^2}{(4\pi)^2} \frac{1}{\bar\epsilon} \left(  -\frac{1}{3} q^2 \right).
\end{eqnarray}
As a result, the \MSbar\ counter-terms $\delta_1$ and $\delta M^2$ are:
\begin{eqnarray}
\delta_1 &=& \frac{N_c \alpha_s}{4\pi} \frac{10}{3} \frac{1}{\bar\epsilon} + \frac{G_T^2 }{(4\pi)^2} \frac{1}{3} \frac{1}{\bar\epsilon}\\
\delta M^2 &=& \frac{N_c \alpha_s}{4\pi} \frac{3M^2}{\bar\epsilon}.
\end{eqnarray}
It can be shown that the $p^\mu p^\nu$ part of the two-point function has the same UV divergence, and that the value of $\delta_1$ above is sufficient to make the entire two-point function finite, as is required by the form of~(\ref{eq:counterterm:gluonfield}).

%
\paragraph{Derivative of two-point function:}
The shift in the residue $\delta R_1$ is given by the derivative of the two-point function (including the counter-terms) evaluated on-shell,
\begin{eqnarray}
%
%
\delta R_1 = 
\left. \frac{d (\Pi_0 + \Pi_\text{CT} )}{dq^2} \right|_{M^2}  &=& \frac{N_c \alpha_s}{4\pi} \Big[\frac{4}{3} \log\frac{\mu^2}{M^2} + \frac{32}{9} - \frac{2}{\bar\epsilon} \Big] 
+ \frac{G_T^2 }{(4\pi)^2} \left[ -\frac{2}{9} - \frac{1}{3} \log\frac{\mu^2}{-M^2} \right].~
\end{eqnarray}
In this expression, the UV diverges cancel by construction leaving behind a purely soft divergence and finite terms.




%
%

\subsection{Vertex corrections} 		
\label{section:triangle}

At NLO in $\alpha_s$, three triangle loop diagrams correct the three-point function. We also include the $\mathcal(g_{L,R}^2)$ correction from a virtual KK gluon. To simplify the Passarino-Veltman decomposition of the triangle loop integrals, we use the \emph{Mathematica} package \emph{Package~X}~\cite{Patel:2015tea}. We calculate the scalar $C_0$ functions by hand in Appendix~\ref{section:handloop}. We follow the notation of \cite{Ellis:2011cr} for the Passarino-Veltman decomposition.


\begin{align*}
\raisebox{-.0\height}{\includegraphics{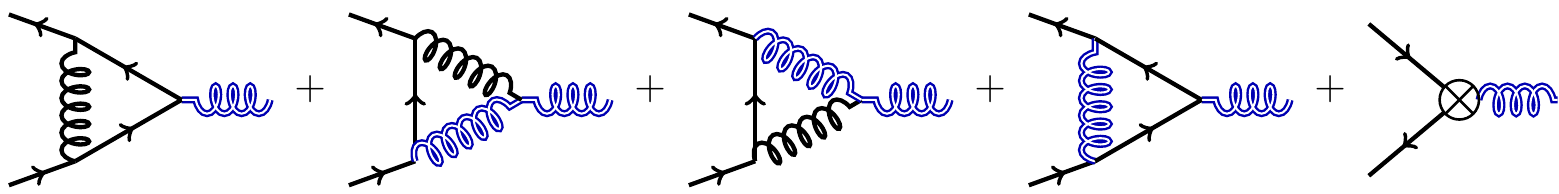}}
\end{align*}
We label these diagrams ``A," ``B," ``C,'' and ``D." Each of these may include UV and IR divergences. We absorb the UV divergences by renormalizing the coupling of the KK gluon to fermions. The IR divergences of the vertex corrections cancel those from the gluon emission process. 


%
\paragraph{Diagram A}
\begin{eqnarray}
i\mathcal M_A & = & \int\! \frac{d^d k}{(2\pi)^d} \bar v_2 (ig_s \gamma^\alpha T^b) \frac{i(\slashed k +\slashed q)}{(k+q)^2} \left(i \gamma^\mu \left[ g_V + g_A \gamma^5 \right]  T^a \right) \frac{i\slashed k}{k^2} (ig_s \gamma^\beta T^b) u_1 \frac{-i \eta_{\alpha\beta}}{(k+p_1)^2} \epsilon_{3\mu}^\star 
\\
&=& g_s^2 (-T^b T^a T^b) \int\! \frac{d^d k}{(2\pi)^d} \frac{v_2 N^\mu u_1 \epsilon_{3\mu}^\star}{k^2 (k+p_1)^2 (k+q)^2},
\end{eqnarray}
where \(q = p_1 + p_2\) and  $N^\mu$ is defined as:
\begin{eqnarray}
N^\mu &=& \gamma^\beta (\slashed k + \slashed q) \gamma^\mu \left[ g_V + g_A \gamma_5 \right] \slashed k \gamma_\beta \\
	&=& \left( (2-2\epsilon)(2 \slashed k k^\mu - k^2 \gamma^\mu) + 4 \slashed k p_2^\mu -4 p_2\cdot k \gamma^\mu -4\epsilon p_1^\mu \slashed k + 4 \epsilon \gamma^\mu p_1\cdot k \right) \left[ g_V + g_A \gamma_5 \right].
\end{eqnarray}
The loop integral can be evaluated using Passarino-Veltman (PV) scalar functions~\cite{Passarino:1978jh}. To separate the UV from the IR divergence, notice that UV divergence requires powers of $k^2$ (or higher). In the unitary gauge we would have powers of $k^4$ from the propagator, but in the Feynman gauge our only UV divergence is from:
\begin{equation}
\int\! \frac{d^d k}{(2\pi)^d} \frac{\gamma^\beta \slashed k \gamma^\mu \slashed k \gamma_\beta}{D_1 \cdot D_2 \cdot D_3} \rightarrow \int\! \frac{d^d k}{(2\pi)^d} \frac{(2)(2 \slashed k k^\mu - k^2 \gamma^\mu)}{D_1 \cdot D_2 \cdot D_3} \ .
\end{equation}
This can be rearranged using the PV tensor function $C^{\mu\nu}$, which has a UV divergence of $\frac{\eta^{\mu\nu}}{4 \epsilon}$.
\begin{equation}
\int\! \frac{d^d k}{(2\pi)^d} \frac{(2-2\epsilon)(2 \slashed k k^\mu - k^2 \gamma^\mu)}{D_1 \cdot D_2 \cdot D_3} \rightarrow \frac{i}{(4\pi)^2} 2\cdot (2\gamma_\alpha \frac{\eta^{\alpha\mu}}{4\epsilon} -  \gamma^\mu \frac{\eta_{\alpha\beta}\eta^{\alpha\beta}}{4\epsilon}) = \gamma^\mu \frac{-1}{\epsilon} + \text{finite}.
\end{equation}
Using \(T^b T^a T^b = (C_F -N_c/2) T^a\), 
\begin{eqnarray}
i\mathcal M_A &=& \frac{i \alpha_s}{4\pi}  \bar v_2 \gamma^\mu T^a \left[ g_V + g_A \gamma_5 \right] u_1 \epsilon_{3\mu}^\star \left(\frac{N_c}{2} - C_F\right) 
\nonumber \\ & & \times
\Big[ 2\Big(\frac{1}{\bar \epsilon^2} + \frac{\log(-\mu^2/q^2)}{\bar\epsilon}\Big) +\frac{3}{\bar\epsilon} +3 \log \frac{-\mu^2}{q^2} + \Big(\log \frac{-\mu^2}{q^2}\Big)^2 + 8 -\frac{\pi^2}{6} \Big].
 \label{eq:amplitude:A}
\end{eqnarray}
All of the terms proportional to $p_1^\mu$ or $p_2^\mu$ also multiply $\slashed p_1$ or $\slashed p_2$, which are proportional to the quark masses and vanish. 

It is useful to separate~(\ref{eq:amplitude:A}) into UV-finite ($\mathcal M^{IR}$) and UV-divergent ($\mathcal M^{UV}$) parts:
\begin{eqnarray}
i\mathcal M_A^{IR} &=& i\mathcal M_{LO} \frac{\alpha_s}{4\pi} \Big(\frac{N_c}{2} - C_F \Big) \Big[ 2\Big(\frac{1}{\bar \epsilon^2} + \frac{\log(-\mu^2/q^2)}{\bar\epsilon}\Big) +\frac{4}{\bar\epsilon}+3 \log \frac{-\mu^2}{q^2} + \Big(\log \frac{-\mu^2}{q^2}\Big)^2 + 8 -\frac{\pi^2}{6} \Big] \nonumber\\  &&\\	\label{eq:ampA}
i\mathcal M_A^{UV} &=& i\mathcal M_{LO} \frac{\alpha_s}{4\pi} \Big(\frac{N_c}{2} - C_F \Big) \Big\{ -\frac{1}{\bar\epsilon} \Big\}.
\end{eqnarray}

%
\paragraph{Diagram B}
\begin{eqnarray}
i\mathcal M_B & = & \int\! \frac{d^d k}{(2\pi)^d} \bar v_2 (i g_s \gamma^B T^b) \frac{i (\slashed k +\slashed p_1)}{(k+p_1)^2} \left(i \gamma^A \left[ g_V + g_A\gamma_5 \right] T^a \right) u_1 \frac{-i\eta_{A \alpha}}{k^2-M^2} \frac{-i\eta_{B\beta}}{(k+q)^2} \nonumber \\
	& & \times g_s f^{abc} \Big[ \eta^{\alpha\beta} (-k -k -q)^\mu +\eta^{\beta\mu}(k+q +q)^\alpha + \eta^{\mu\alpha}(-q +k)^\beta \Big] \epsilon_{3\mu}^\star \\
&=& i g_s^2 T^c T^b f^{abc} \int\! \frac{d^d k}{(2\pi)^d} \frac{\bar v_2 N^\mu u_1 \epsilon_{3\mu}^\star }{(k^2 - M^2) (k+p_1)^2 (k+q)^2}
\end{eqnarray}
where $N^\mu$ is:
\begin{eqnarray}
N^\mu &=& \gamma_\beta (\slashed k \gamma_\alpha + 2 p_{1\alpha})\Big[\eta^{\alpha\beta}(-2k - q)^\mu + \eta^{\beta\mu}(k+2q)^\alpha + \eta^{\mu\alpha} (k-q)^\beta \Big]\left[ g_V + g_A\gamma_5 \right].
\end{eqnarray}
After completing the loop integral, the amplitude is:
\begin{equation}
i\mathcal M_B = g_s^2  \frac{i N_c }{2(4\pi)^2} \bar v_2 \gamma^\mu  \left[ g_V + g_A\gamma_5 \right] T^a u_1 \epsilon_{3\mu}^\star \Big[ -1 -\frac{1}{\bar\epsilon^2} -\frac{\log (\mu^2/M^2)}{\bar\epsilon}  - \frac{1}{2} \Big( \log \frac{\mu^2}{M^2}\Big)^2 + \frac{1}{\bar\epsilon} +  \log \frac{\mu^2}{M^2} \Big].
\end{equation}
The UV divergence of this amplitude arises from:
\begin{eqnarray}
2 k^2 \gamma^\mu + (4 - 4\epsilon) \slashed k k^\mu &\rightarrow& 2\eta_{\alpha \beta} C^{\alpha \beta} + 4 \gamma_\alpha C^{\alpha\beta} \\ 
&\rightarrow & \frac{i}{(4\pi)^2} \frac{8+4}{4 \epsilon} \gamma^{\mu} + \text{finite}.
\end{eqnarray}
As before, we separate the UV divergence from the rest of the amplitude:
\begin{eqnarray}
i\mathcal M_B^{IR} &=& \frac{\alpha_s}{4\pi} \frac{N_c}{2} i \mathcal M_{LO} \Big[ -\frac{1}{\bar\epsilon^2} -\frac{\log (\mu^2/M^2)}{\bar\epsilon} - \frac{2}{\bar\epsilon} +  \log \frac{\mu^2}{M^2} -\frac{1}{2} \Big( \log \frac{\mu^2}{M^2} \Big)^2 - 1 -\frac{\pi^2}{12} \Big] \\ 	\label{eq:ampB}
i\mathcal M_B^{UV} &=& \frac{\alpha_s}{4\pi} \frac{N_c}{2} i \mathcal M_{LO} \Big\{ \frac{3}{\bar\epsilon} \Big\}.
\end{eqnarray}

\paragraph{Diagram C}
In the Feynman gauge, the numerator structure can be made identical to that of Diagram~B by commuting the 
$\left[ g_Q + g_A\gamma_5 \right]$ twice to the right. The difference is the location of $M^2$ in the denominator:
\begin{equation}
i\mathcal M_B  = i g_s^2 T^c T^b f^{abc} \int\! \frac{d^d k}{(2\pi)^d} \frac{\bar v_2 N^\mu u_1 \epsilon_{3\mu}^\star }{(k^2 ) (k+p_1)^2 ((k+q)^2-M^2)}.
\end{equation}
If we perform the shift $k \rightarrow k-q$ and then an inversion in $k$, we can make the denominators match. 
Applying this transformation to the numerator $N^\mu$ of Diagram~B has the effect of switching $p_1 \leftrightarrow p_2$. 
This would change the amplitude, if not for $p_1^2=p_2^2=0$; the only nonzero invariant 
$p_1 \cdot p_2 = q^2/2$ remains unchanged by the $p_1 \leftrightarrow p_2$ transformation.
\begin{equation}
i\mathcal M_C = i\mathcal M_B.
\end{equation}

\paragraph{Diagram D}
This amplitude is very similar to the one in Diagram~A, with small changes in the numerator and with one massive propagator.
\begin{eqnarray}
i\mathcal M_D &=& \int\!\frac{d^d k}{(2\pi)^d} \bar v_2 \left[ i \gamma^\alpha (P_L g_L + P_R g_R) T^b \right] \frac{ i (-\slashed k - \slashed q)}{(k+q)^2} \left[ i \gamma^\mu  (P_L g_L + P_R g_R)  T^a \right] \frac{-i \slashed k}{k^2} \nonumber\\&&\ \times \left[i \gamma_\alpha (P_L g_L + P_R g_R)  T^b \right]  \frac{-i}{(k+p_1)^2 - M^2} u_1 \epsilon_{3\mu}^\star \\
&=& \left( \frac{N_c}{2} - C_F \right) \frac{i }{(4\pi)^2} \left[ -\frac{1}{\bar\epsilon} - \log\frac{\mu^2}{M^2} + 6 + 5 i \pi - \frac{8\pi^2}{3} + 8\, \text{Li}_2(2) \right] \bar v_2 \gamma^\mu (P_L g_L^3 + P_R g_R^3 ) T^a u_1 \epsilon_{3\mu}^\star. \nonumber\\
\end{eqnarray}
In analogy with $\mathcal M_\text{LO}$, we define $\mathcal M_\text{KK}'$ as follows:
\begin{equation}
\mathcal M_\text{KK}' = \bar v_2 i \gamma^\mu (P_L g_L^3 + P_R g_R^3 ) T^a u_1 \epsilon_{3\mu}^\star.
\end{equation}
The UV divergent and finite parts of $\mathcal M_D$ are:
\begin{eqnarray}
\mathcal M_D^{\text{IR}} &=& \left( \frac{N_c}{2} - C_F \right) \frac{i \mathcal M_\text{KK}'}{(4\pi)^2} \left[ - \log\frac{\mu^2}{M^2} + 6 + 5 i \pi - \frac{8\pi^2}{3} + 8\, \text{Li}_2(2) \right]   \\
\mathcal M_D^{\text{UV}} &=& \left( \frac{N_c}{2} - C_F \right) \frac{i \mathcal M_\text{KK}'}{(4\pi)^2} \left[ - \frac{1}{\bar\epsilon} \right].  
\end{eqnarray}

%
%
\subsubsection{Gluon Mixing Amplitude}
In addition to the self-energy and vertex corrections, there is a diagram in which the QCD vertex is attached to a bubble of quarks
which mixes the gluon and the KK gluon:
\begin{align*}
\sum_{q'}~
\left( 
\raisebox{-.44\height}{\includegraphics{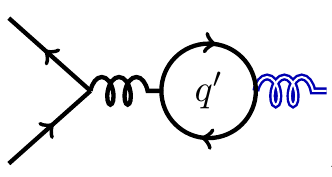}}
\right).
\end{align*}
where the sum over $q'$ includes all flavors of quark in the loop.
While it appears to be a mixed self-energy, it vanishes when the gluon is on-shell, and formally contributes to the renormalization
of the coupling of the SM quarks to the KK gluon.
Unlike the leading-order amplitude, which was proportional to $\left[ g_V^{(q)} + g_A^{(q)}\gamma_5 \right]$, this amplitude is 
proportional to a sum over $q'=d,u,s,c,b,t$:
\begin{equation}
i \mathcal M_M(q\bar q \rightarrow G) \sim g_s^2 \sum_{q'} \left[ g_V^{(q')} + g_A^{(q')}\gamma_5 \right].
\end{equation}
The gluon mixing two-point function $\Pi^{\alpha\beta}_{M}(q^2)$ is IR finite but UV divergent,
and is contracted with a QCD vertex,
\begin{equation}
i\mathcal M_M = \left(\bar v_2 ig_s \mu^\epsilon \gamma^\beta T^a u_1 \right) \frac{-i \eta_{\beta\alpha} }{q^2} i\Pi_{M}^{\alpha \mu}(q^2) \delta^{ab} \epsilon_{3\mu}^{\star b}.
\end{equation}
We may drop any terms proportional to the gluon momentum in 
$\Pi_M^{\alpha\mu}$ proportional to $q^\mu$, because $q^\mu \epsilon_{3\mu}^\star = 0$.
\begin{eqnarray}
i\Pi_{M}^{\alpha\mu} \delta^{ab} &=& \int\!\frac{d^d k}{(2\pi)^d} \frac{(-1) \text{Tr}\!\left[ i g_s \mu^\epsilon T^a \gamma^\alpha \cdot i\slashed k \cdot i \gamma^\mu [g_V + g_A \gamma_5] T^b \cdot i (\slashed k + \slashed q) \right]}{k^2 (k+q)^2} \\
i\Pi_{M}^{\alpha\mu} &=& - \frac{g_s g_V }{2} \int_0^1\! dx \int\!\frac{d^d \ell}{(2\pi)^d} \left[ \frac{1}{\ell^2 - \Delta} \right]^2 \mathbb{N}^{\alpha\mu},
\end{eqnarray}
where $\Delta = - q^2 x (1-x)$ and 
\begin{equation}
\mathbb{N}^{\alpha\mu} = 4 \left[ 2 \frac{\ell^2}{d} - \ell^2 + q^2( x - x^2) \right] \eta^{\alpha\mu} + q^\alpha q^\mu (\ldots)\ .
\end{equation}
The $g_A$ term is proportional to $4i \epsilon_{\alpha\mu\rho\sigma} q^\rho q^\sigma = 0$. 
After discarding the $q^\alpha q^\mu$ part of the two point function and summing over loop quark flavors, the remainder is:
\begin{eqnarray}
i\Pi_{M} \eta^{\alpha\mu} &=&\eta^{\alpha\mu} \sum_{(q')} g_s (2 g_V^{(q')}) \frac{ i \Gamma(\epsilon) }{(4\pi)^2} \int_0^1 \! dx \left( \frac{4\pi\mu}{\Delta} \right)^\epsilon (-2 q^2)(x-x^2) \\
\mu^\epsilon \Pi_{M}  &=& g_s \sum_{(q')} \left( g_L^{(q')} + g_R^{(q')} \right) \frac{ \Gamma(\epsilon) }{ (4\pi)^2} \left( \frac{ 4\pi \mu^2}{ - q^2} \right)^\epsilon (-2q^2) \frac{ \Gamma(2-\epsilon) \Gamma(2-\epsilon) }{\Gamma(4-2\epsilon)} .
\end{eqnarray}
We define $G_Q' $ and $i \mathcal M_{\text{QCD}}'$ as follows:
\begin{align}
G_Q' \equiv \sum_{(q')} \left( g_L^{(q')} + g_R^{(q')} \right)
&&
i \mathcal M_{\text{QCD}}' \equiv\bar v_2 i  \gamma^\mu G_Q' T^a u_1 \epsilon_{3\mu}^\star  .
\end{align}
Now the amplitude $i\mathcal M_D$ may be written in a compact form:
\begin{align}
i \mathcal M_M = g_s \mu^\epsilon \frac{i \mathcal M_{\text{QCD}}'}{G_Q'} \frac{\Pi_M (q^2)}{q^2} 
= \left(i \mathcal M_{\text{QCD}}' \right) \frac{\alpha_s}{4\pi} \left[ \frac{-1}{3} \left( \frac{1}{\bar\epsilon} + \log\frac{\mu^2}{-q^2} \right) - \frac{5}{9} \right] ,
\end{align}
which we split into finite and UV-divergent parts:
\begin{eqnarray}
i \mathcal M_M^\text{finite} &=& \left(i \mathcal M_{\text{QCD}}' \right) \frac{\alpha_s}{4\pi}  \left[ -\frac{1}{3}  \log\frac{\mu^2}{-q^2}  - \frac{5}{9} \right], \\
i \mathcal M_M^\text{UV} &=& \left(i \mathcal M_{\text{QCD}}' \right) \frac{\alpha_s}{4\pi}  \left( -\frac{1}{3} \frac{1}{\bar\epsilon} \right).
\end{eqnarray}

\subsubsection{Coupling Renormalization}

The counter-terms for the couplings of the quarks to the KK gluon take the form:
\begin{align}
\raisebox{-.48\height}{\includegraphics{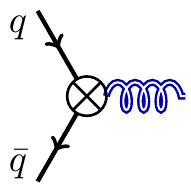}}
\;
=
\;
i \gamma^\mu \left[ \delta_L^{(q)} g_L^{(q)} P_L + \delta_R^{(q)} g_R^{(q)} P_R \right] T_a .
\end{align}
The values of $\delta_L$ and $\delta_R$ are determined in the \MSbar\ scheme by the UV-divergent terms:
\begin{eqnarray}
i\mathcal M_{UV} &=& i\mathcal M_{\text{CT}} + i\mathcal M_A^{UV} + i\mathcal M_B^{UV} + i\mathcal M_C^{UV} + i\mathcal M_D^{UV} + i\mathcal M_M^{UV} = 0\ .
\end{eqnarray}
The scalings $Z_L^{q}$ and $Z_R^{q}$ are chosen so that the amplitude is UV-finite:
\begin{eqnarray}
\left[ \delta_L^{(q)} g_L^{(q)} P_L + \delta_R^{(q)} g_R^{(q)} P_R \right] &=&\frac{-\alpha_s}{4\pi} \frac{1}{\bar\epsilon} \bigg[ \left( g_L^{(q)} P_L + g_R^{(q)} P_R \right) \left[ C_F + \frac{5 N_c}{2} \right] 
 + \left( P_L + P_R \right) \left( -\frac{G_Q'}{3} \right) \bigg], \nonumber\\
\end{eqnarray}
such that
\begin{eqnarray}
\delta_L^{(q)} &=& \frac{\alpha_s}{4\pi} \frac{1}{\bar\epsilon} \left[ - C_F - \frac{5N_c}{2} + \frac{1}{3} \frac{G_Q'  }{g_L^{(q)} } \right] + \frac{g_L^{(q)2} }{(4\pi)^2} \frac{1}{\bar\epsilon}  \left( \frac{N_c}{2} - C_F \right) ,\\
\delta_R^{(q)} &=& \frac{\alpha_s}{4\pi} \frac{1}{\bar\epsilon} \left[ - C_F - \frac{5N_c}{2} + \frac{1}{3}  \frac{G_Q' }{g_R^{(q)} } \right] + \frac{g_R^{(q)2}}{(4\pi)^2} \frac{1}{\bar\epsilon} \left( \frac{N_c}{2} - C_F \right)  . 
\end{eqnarray}
We use $Z_1$ and $Z_2$ from the gluon and quark two-point functions to find the $Z_L^q$ and $Z_R^q$ that correspond to these counter-terms.
\begin{eqnarray}
\delta_{L,R}^q &=& Z_{L,R}^q Z_Q \sqrt{Z_1} - 1\, = \, \delta Z_{L,R}^q + \delta_Q + \frac{1}{2} \delta_1 \\ 
%
\delta Z_{L,R}^q & = & \frac{\alpha_s}{4\pi} \frac{1}{\bar\epsilon} \Big(-C_F -\frac{5 N_c}{2}  +  \frac{G_Q' }{3 g_{L,R}^{(q)} } + C_F - \frac{1}{2} \frac{10 N_c}{3} \Big) + \frac{g_{L,R}^{(q)2}}{(4\pi)^2} \frac{1}{\bar\epsilon} \left( C_F + \frac{N_c}{2} - C_F \right)   \\
Z_{L,R}^{(q)} &=&1 +\frac{\alpha_s}{4\pi} \frac{1}{\bar\epsilon} \Big( -\frac{25 }{6} N_c + \frac{G_Q' }{3 g_{L,R}^{(q)} }  \Big) + \frac{g_{L,R}^{(q)2}}{(4\pi)^2} \frac{1}{\bar\epsilon} \left( \frac{N_c}{2} \right)  .
\end{eqnarray}
As expected from the Ward identity, the divergences proportional to $C_F$ cancel between counter-terms.

\subsection{Virtual Correction to the Cross Section}	
\label{section:virtual:cross}

We now have all of the ingredients necessary to assemble the virtual correction to the cross section. For the terms in $\sigma_\text{virt}$ not proportional to $\sigma_0$ we define $\sigma_\text{KK}$ and $\sigma_\text{QCD}$ as follows:
\begin{align}
\sigma_\text{KK} \;\equiv\; \frac{\pi C_F}{2 N_c} (g_L^4 + g_R^4) (1-\epsilon), &&
\sigma_\text{QCD}^{(q)} \; \equiv\;  G_Q'  \frac{\pi C_F}{2 N_c} \left( g_L^{(q)} + g_R^{(q)} \right) .
\end{align}
Assembling the various pieces into Eq.~(\ref{eq:amplitude:LSZ}) leads to the final result,
\begin{eqnarray}
\sigma_{\text{virt}}^{(q)} &=& \delta(s-M^2) \Bigg( \frac{\alpha_s \sigma_0^{(q)}}{4\pi}  \bigg[ C_F \Big\{ -\frac{4}{\bar\epsilon^2} - \frac{4 \log(\mu^2/M^2)}{\bar\epsilon} - \frac{2}{\epsilon} - 10 +\frac{7}{3} \pi^2 - 2 \log\frac{\mu^2}{M^2} -2 \Big(\log\frac{\mu^2}{M^2} \Big)^2 \Big\} \nonumber\\&&\
+ N_c \Big\{ -\frac{2}{\bar\epsilon} + \frac{104}{9} -\frac{4}{3} \pi^2 + \frac{19}{3} \log\frac{\mu^2}{M^2} \Big\}  \bigg] 
+ \frac{\alpha_s \sigma_\text{QCD}^{(q)}}{4\pi} \left[ -\frac{2}{3} \log\frac{\mu^2}{M^2} - \frac{10}{9} \right]  \\&&\
+ \frac{\sigma_\text{KK}^{(q)}}{(4\pi)^2} \left[ C_F\left( \frac{16 \pi^2}{3} - 11 - 16\, \text{Re}[ \text{Li}_2(2) ] \right) + N_c \left( -\log\frac{\mu^2}{M^2} + 6 - \frac{8\pi^2}{3} + 8\, \text{Re}[ \text{Li}_2(2) ] \right) \right] 
\Bigg), \nonumber
\end{eqnarray}
%
%
%
which we separate into an IR divergent piece and finite remainder, to make it easier to cancel with the real correction:
\begin{eqnarray}
\sigma_{\text{virt}}^\text{soft} &=&\frac{\alpha_s}{4\pi} \sigma_0^{(q)} \delta(s-M^2)  \Big[ C_F \Big\{ -\frac{4}{\bar\epsilon^2} - \frac{4 \log(\mu^2/M^2)}{\bar\epsilon} - \frac{2}{\epsilon} \Big\} - N_c \frac{2}{\bar\epsilon} \Big] \\	\label{eq:virtual:soft}
\sigma_{\text{virt}}^\text{finite} &=&  \delta(s-M^2) \Bigg( 
 \frac{\alpha_s \sigma_0^{(q)}}{4\pi} \bigg[ C_F \Big\{  - 10 +\frac{7}{3} \pi^2 - 2 \log\frac{\mu^2}{M^2} -2 \Big(\log\frac{\mu^2}{M^2} \Big)^2 \Big\} 
 \\ &&\
+ N_c \Big\{ + \frac{140}{9} -\frac{4}{3} \pi^2 + \frac{19}{3} \log\frac{\mu^2}{M^2} \Big\} \bigg] 
+ \frac{\sigma_\text{KK}^{(q)}}{(4\pi)^2} \bigg[ C_F\left( \frac{16 \pi^2}{3} - 11 - 16\, \text{Re}[ \text{Li}_2(2) ] \right) \nonumber\\&&\
 + N_c \left( -\log\frac{\mu^2}{M^2} + 6 - \frac{8\pi^2}{3} + 8\, \text{Re}[ \text{Li}_2(2) ] \right) \bigg] 
+ \frac{\alpha_s \sigma_\text{QCD}^{(q)} }{4\pi} \left[ -\frac{2}{3} \log\frac{\mu^2}{M^2} - \frac{10}{9} \right]     \nonumber
\Bigg).
 \label{eq:virtual:finite}
\end{eqnarray}


\section{Real Corrections}		
\label{section:real}

In this section we compute the ${\mathcal O}(\alpha_s)$ corrections from the radiative processes $q\bar q \rightarrow g G$, $q g \rightarrow q G$,
and $\bar q g \rightarrow \bar q G$ (the latter two are NLL for an initial state top quark).
These contributions contain collinear divergences which have been absorbed into the definition of the PDFs and are removed by \MSbar\ counter-terms,
and the $q\bar q \rightarrow g G$ process additionally contains soft divergences which cancel with those in the virtual corrections.

We describe the $2 \rightarrow 2$ scattering kinematics with Mandelstam variables $s\equiv(p_1+p_2)^2=(p_3+p_4)^2$, 
$t\equiv(p_3-p_1)^2=(p_4-p_2)^2$, $u\equiv(p_4-p_1)^2=(p_3-p_2)^2$, only two of which are independent because $s+t+u = M^2$.  
The IR and collinear divergences are regulated by integrating over $d$-dimensional phase space:
\begin{equation}
\sigma=\frac{1}{2s}\int\! \frac{d^{d-1} p_3}{(2\pi)^{d-1} 2 p_3^0} \frac{d^{d-1} p_4}{(2\pi)^{d-1} 2 p_4^0} (2\pi)^d \delta^d (p_1+p_2-p_3-p_4) \left|\mathcal M\right|^2.
\label{eq:sigma:qEmission}
\end{equation}
For gluons and massless quarks,
\begin{eqnarray}
\frac{1}{2s} \Pi^{(d)}_2 &=& \frac{1}{2s} \int\! \frac{d^{d-1} p_3}{(2\pi)^{d-1} 2 p_3^0} \frac{d^{d-1} p_4}{(2\pi)^{d-1} 2 p_4^0} (2\pi)^d \delta^d (p_1+p_2-p_3-p_4) \\
& = & \frac{s-M^2}{32 \pi^2 s^2} \Big(\frac{4 \pi s}{(s-M^2)^2}\Big)^\epsilon \frac{\Gamma(1-\epsilon)}{\Gamma(1-2\epsilon)} \int\! d\Theta,
\end{eqnarray}
where we define $\int\! d\Theta$:
\begin{equation}
\int\! d\Theta = \int_0^\pi\!d\theta (\sin\theta)^{1-2\epsilon} \int_0^\pi\!d\phi (\sin\phi)^{-2\epsilon}.
\end{equation}
The cross section is
\begin{eqnarray}
\sigma & = & \frac{s-M^2}{32\pi^2s^2} \Big( \frac{4\pi}{s} \Big)^\epsilon \Big[\frac{1}{1-\tau}\Big]^{2\epsilon} \frac{\Gamma(1-\epsilon)}{\Gamma(1-2\epsilon)} \int\! d\Theta \left| \mathcal M \right|^2.
\label{eq:sigma:gEmission}
\end{eqnarray}
The necessary integrals are evaluated in Ref.~\cite{Beenakker:1988bq} and tabulated in terms of $t=(M^2-s)(1+\cos\theta)/2$ and $u=(M^2-s)(1-\cos\theta)/2$
in Appendix~\ref{section:integraltables}.


\subsection{$q\bar q \rightarrow g G$}

The radiative process $q\bar q \rightarrow g G$ contain both soft and collinear singularities, which are regulated by the $d$ dimensional phase space
as in Eq.~(\ref{eq:sigma:gEmission}).

\paragraph{Amplitude}
\label{section:emission:gluon}

We assign $p_1$ to the incoming quark, $p_2$ to the incoming anti-quark, $(p_3,\mu)$ to the KK gluon and $(p_4,\nu)$ to the massless gluon.
There are three Feynman diagrams,
\begin{align*}
\raisebox{-.35\height}{\includegraphics{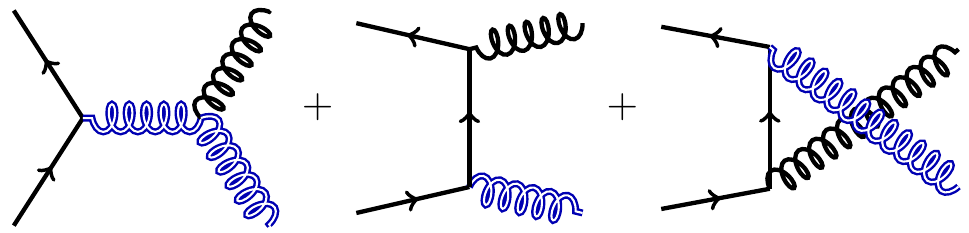}}\ ,
\end{align*}
whose amplitudes we label as $\mathcal M_S$, $\mathcal M_T$, and $\mathcal M_U$, respectively.  They are:
\begin{eqnarray}
i\mathcal M_S &=& \frac{ g_s T^c f^{abc}}{s-M^2}  \bar v_2 \Big( \eta^{\mu\nu}(\slashed p_4 -\slashed p_3) -2\gamma^\nu p_4^\mu + 2\gamma^\mu p_3^\nu \Big)  [g_V + g_A \gamma_5] u_1 \epsilon_{4\nu}^{b\star} \epsilon_{3\mu}^{a\star}.\\
i\mathcal M_T &=& \bar v_2 i g_s T^b_{jk} \gamma^\nu \frac{i(\slashed p_1 - \slashed p_3)}{t} i T^a_{ki} \gamma^\mu [g_V + g_A \gamma_5] u_1 \epsilon_{4\nu}^{b\star} \epsilon_{3\mu}^{a\star} \\
i\mathcal M_U &=& \bar v_2 i T^a_{jk} \gamma^\mu  [g_V + g_A \gamma_5] \frac{i(\slashed p_1 - \slashed p_4)}{u} i g_s T^b_{ki} \gamma^\nu u_1 \epsilon_{4\nu}^{b\star} \epsilon_{3\mu}^{a\star}.
\end{eqnarray}
We have used the fact that $p_4 \cdot \epsilon_4 = 0$ to simplify $\mathcal M_S$ and retain only physical polarizations.  Squaring the net amplitude and 
summing/averaging over spins and colors, the cross section is evaluated via Eq.~(\ref{eq:sigma:gEmission})
using angular integrals found in Appendix~\ref{section:integraltables}.  The resulting cross section contains soft 
(including soft {\em and} collinear) and collinear divergences.

\paragraph{Soft divergences}

The soft divergences arise from factors of $(1-\tau)^{-1}$ in the matrix elements, and
are regulated by the factor of $(1-\tau)^{2\epsilon}$ in~Eq.~(\ref{eq:sigma:gEmission}).
It is convenient to
define a ``plus distribution" with the fractional power of $(1-\tau)^{2\epsilon}$ from the phase space integral. 
The integral over the incoming parton momentum fractions (see Section~\ref{sec:results}) thus contains:
\begin{eqnarray}
\int_{\tau_0}^1\! d\tau \Big(\frac{1}{1-\tau} \Big)^{1+2\epsilon} &=& \int_{\tau_0}^1\! d\tau \bigg\{ \Big(\frac{1}{1-\tau} \Big)^{1+2\epsilon} -  \delta(1-\tau)\int_{\tau_0}^1 \frac{dz}{(1-z)^{1+2\epsilon}} - \frac{\delta(1-\tau)}{2\epsilon} \bigg\} \\
& = &   \int_{\tau_0}^1\! d\tau \bigg\{ \Big[\frac{1}{(1-\tau)^{1+2\epsilon}} \Big]_+ -\frac{1}{2\epsilon} \delta(1-\tau) \bigg\},
\end{eqnarray}
in terms of the plus distribution defined as:
\begin{equation}
\int_{\tau_0}^1\! d\tau\ f(\tau) \Big[\frac{1}{1-\tau} \Big]_+ \equiv \int_{\tau_0}^1\! d\tau\ \frac{f(\tau) - f(1)}{1-\tau}.
\end{equation}
The soft divergences are thus exposed:
\begin{equation}
\Big[\frac{1}{1-\tau}\Big]^{1+2\epsilon} = \; - \frac{1}{2\epsilon} \delta(1-\tau) + \Big[ \frac{1}{(1-\tau)}\Big]_+ - 2\epsilon \Big[ \frac{\log(1-\tau)}{(1-\tau)} \Big]_+ .
\end{equation}
In the process $q\bar q \rightarrow g G$, the IR divergent terms are:
\begin{equation}
\sigma_{\text{real}}^{\text{soft}} =  \alpha_s (g_V^2 + g_A^2) \frac{ \delta(1-\tau)}{N_c^2 M^2} \bigg( \frac{N_c^2 C_F}{2} \frac{1}{\bar\epsilon}  + C_F^2 N_c \Big\{ \frac{1}{\bar\epsilon^2} + \frac{\log(\mu^2/s)}{\bar\epsilon}  + \frac{1}{2} \frac{1}{\bar\epsilon} \Big\} \bigg).	
\label{eq:real:soft}
\end{equation}
where the $1 / \bar \epsilon^2$ terms represent the overlapping soft and collinear singularities.

\paragraph{Collinear divergences and PDF Counter-term}

In \MSbar\ the correction to the hard scattering matrix element for the $q \bar q$ initial state is given by:
\begin{eqnarray}
H_{q\bar q}^{(1)} & = & \sigma_{q\bar q}^{\text{virt}} + \sigma_{q\bar q}^{\text{real}} 
- \frac{\alpha_s}{\pi} \Big[ \phi_{q \rightarrow q}^{(1)} \otimes \sigma_\text{LO} + \phi_{\bar q \rightarrow \bar q}^{(1)} \otimes  \sigma_\text{LO} \Big] \\
& \equiv & \sigma_{q\bar q}^{\text{virt}} + \sigma_{q\bar q}^{\text{real}} + \sigma_{q\bar q}^{CT}.
\end{eqnarray}
The splitting functions $q \rightarrow q$ and $\bar q \rightarrow \bar q$ are given by:
\begin{equation}
\phi_{q \rightarrow q} = \phi_{\bar q \rightarrow \bar q} = -\frac{1}{2\bar\epsilon} C_F \Big[\frac{1+z^2}{1-z}\Big]_+
\end{equation}
where $z$ is the fraction of momentum of the parent carried by the daughter.
$ \sigma_{q\bar q}^{CT}$ is the leading order cross section convolved with the splitting function,
\begin{eqnarray}
\sigma_{q\bar q}^{CT} &=& \frac{\alpha_s}{\pi} \frac{C_F}{\bar\epsilon} \int_0^1\! \text{d}z 
\Big(\frac{1+z^2}{[1-z]_+} + \frac{3}{2} \delta(1-z) \Big) ~ \times ~  \sigma_0 (1-\epsilon)\delta(s z - M^2) \\
 &=& \frac{\alpha_s (g_L^2+g_R^2) N_c C_F^2}{2N_c^2 s}\frac{1-\epsilon}{\bar\epsilon} \Big[\frac{1+\tau^2}{[1-\tau]_+} + \frac{3}{2} \delta(1-\tau) \Big]. 	\label{eq:counterterm:gluon}
\end{eqnarray}
The counter-term contains both collinear and soft divergences. The soft divergence multiplies $\delta(1-\tau)$,
allowing it to be easily combined with the soft/virtual corrections.

\paragraph{Hard Contribution}

Combining the matrix elements for $q\bar q \rightarrow g G$ and the PDF counter-term results in an expression which is finite for
$\tau \neq 1$.  We verify that the remaining terms proportional to $\delta(1-\tau)$ cancel the IR divergences in the virtual
corrections, Eq.~(\ref{eq:virtual:soft}).  What remains is the finite contribution to the hard scattering cross section,
\begin{eqnarray}
\sigma_{\text{real g}}^{\text{finite}}
&=& \frac{\alpha_s }{4\pi}\frac{\sigma_0}{M^2} \bigg\{ \delta(1-\tau) \Big\{ N_c \Big(-2 + 2\lgmus \Big)  
+ C_F \Big( - 6 -\pi^2 - 4\lgmus + 2 \Big[\lgmus\Big]^2 \Big) \Big\} \nonumber\\&&\
+ \Big[\frac{1}{1-\tau} \Big]_+ \Big\{ N_c \Big( \frac{2}{3} \tau^3 -\frac{10}{3} \tau^2 -\frac{10}{3} \tau + 2 \Big)
+ C_F \Big(-4\tau^3 + 6\tau^2 -2 - 8\tau^2 \lgmus \Big) \Big\} \nonumber\\&&\
+ \Big[\frac{\log(1-\tau)}{1-\tau}\Big]_+ 8 C_F \Big(\tau + \tau^3\Big)
+ N_c \Big(2\tau^2 -2 \Big)  \nonumber\\&&\
+ C_F \Big(2 + 2\tau -4\tau^2 + (4\tau^2 - 4\tau ) \lgmus \Big)
\bigg\}.		\label{eq:real:finite}
\end{eqnarray}


\subsection{$q g \rightarrow q G$  (Light Quarks)}
\label{sec:masslessq}

In computing the hard scattering cross section for $q g \rightarrow q G$, we denote by $p_1$ and $(p_2,\nu)$ the incoming quark and gluon respectively; 
$(p_3,\mu)$ and $p_4$ correspond to the KK gluon and the outgoing quark. 
We focus the discussion on $q g \rightarrow q G$ since the
cross section for $\bar q g \rightarrow \bar q G$ is the same as for $q g \rightarrow q G$.
There are three Feynman diagrams we denote by $s$-channel, $t$-channel, and $u$-channel,
\begin{align*}
\raisebox{-.4\height}{\includegraphics{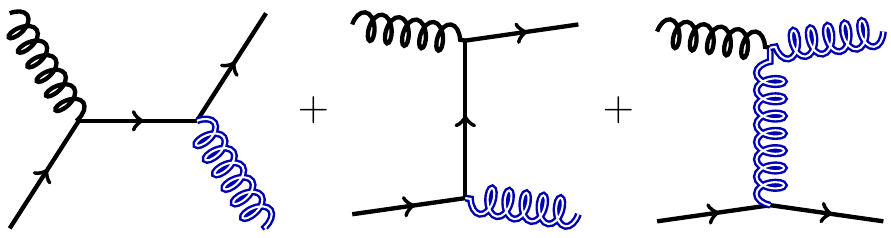}} ~.
\end{align*}
The three amplitudes simplify to
\begin{eqnarray}
\mathcal M_S & =& g_s \mu^\epsilon C_s \bar{u_4} \mathbb{N}^{\mu\nu}_s  u_1\epsilon^\star_{3\mu} \epsilon_{2\nu}	\label{eq:quark:ampS} \\
\mathcal M_T &=& g_s \mu^\epsilon C_t \bar{u_4} \mathbb{N}^{\mu\nu}_t  u_1\epsilon^\star_{3\mu} \epsilon_{2\nu}  \label{eq:quark:ampT} \\
\mathcal M_U &=& g_s \mu^\epsilon C_u \bar{u_4} \mathbb{N}^{\mu\nu}_u u_1\epsilon^\star_{3\mu} \epsilon_{2\nu}  \label{eq:quark:ampU}\, ,
\end{eqnarray}
where we have used the Ward identity to simplify $\mathcal M_U$, and where
\begin{eqnarray}
C_s = -\frac{T^a T^b}{s} 		&\hspace{0.5cm} &	\mathbb{N}^{\mu\nu}_s = \gamma^\mu (\slashed p_1 + \slashed p_2)\gamma^\nu [g_V + g_A \gamma_5] \\
C_t = -\frac{T^b T^a}{t} 		& &	\mathbb{N}^{\mu\nu}_t = \gamma^\nu (\slashed p_1 - \slashed p_3) \gamma^\mu [g_V + g_A \gamma_5] \\
C_u = \frac{i T^c f^{abc} }{u-M^2} 	& & \mathbb{N}^{\mu\nu}_u = \left[ \eta^{\mu\nu} (\slashed p_2 + \slashed p_3 ) - 2 p_2^\mu \gamma^\nu - 2 p_3^\nu \gamma^\mu \right] (g_V + g_A \gamma_5).
\end{eqnarray}
The amplitude is squared and summed/averaged over spins and colors.  The necessary angular integrals to
determine the cross section are tabulated in Appendix~\ref{section:integraltables}.  The result contains no soft singularities, but does include collinear ones.

\paragraph{Collinear Divergences and Counter-term}

The hard scattering matrix element is given by,
\begin{equation}
H_{gq}=\sigma_{gq}^{\text{real}} - \frac{\alpha_s}{\pi} ~\phi^{(1)}_{g \rightarrow \bar{q}} \otimes \sigma_\text{LO} .		 
\label{eq:Hgq:def}
\end{equation}
The subtracted term on the right side of the equation is the counter-term:
\begin{eqnarray}
\sigma^{CT}_{gq}(\tau)
&=& \frac{(g_V^2 + g_A^2) \alpha_s C_F^2}{2 (N_c^2 -1) M^2} \Big(\frac{1}{\bar\epsilon} - 1\Big) (2\tau^3 -2\tau^2 + \tau).
\end{eqnarray}

\paragraph{Hard Contribution}
Combining the counter-term with the real emission diagrams leads to a result which is finite:
\begin{eqnarray}
\sigma_{\text{real q}}^{\text{finite}} &=& \frac{\alpha_s}{4\pi} \frac{\sigma_0}{2 C_F M^2} 
\bigg[ C_F  \bigg\{ \tau + 6\tau^2 - 7\tau^3 + (-2\tau + 4\tau^2 - 4\tau^3) \left( \log\frac{s\mu^2}{(s-M^2)^2}  \right) \bigg\} \nonumber\\&&\
+ N_c \bigg\{ 4 +4\tau + 2\tau^2 - 4\tau^3 + (4\tau^2 + 4\tau) \log\tau \bigg\}  \bigg].
\label{eq:massless:final}
\end{eqnarray}
The expression for $\sigma_{\text{real~$\bar q$}}^{\text{finite}}$ is identical.


\subsection{$t g \rightarrow t G$} 
\label{section:massivequark}

Following the m-ACOT prescription, we retain the heavy quark mass in computing the processes where a top quark fuses
with one or more initial state gluons.  As a result, there is no need
to dimensionally continue the phase space integral, as the collinear singularities are regulated by the presence of the top mass.  The computation proceeds
similarly to the light quark case of Section~\ref{sec:masslessq} in terms of $s$, $t$, and $u$-channel Feynman diagrams, with amplitudes:
\begin{align}
\mathcal M_S = g_s \mu^\epsilon C_s \bar{u_4} \mathbb{N}^{\mu\nu}_s  u_1\epsilon^\star_{3\mu} \epsilon_{2\nu}  &,&
\mathcal M_T = g_s \mu^\epsilon C_t \bar{u_4} \mathbb{N}^{\mu\nu}_t  u_1\epsilon^\star_{3\mu} \epsilon_{2\nu}   &,&
\mathcal M_U = g_s \mu^\epsilon C_u \bar{u_4} \mathbb{N}^{\mu\nu}_u u_1\epsilon^\star_{3\mu} \epsilon_{2\nu}.
\end{align}
In this section,
\begin{eqnarray}
C_s = -\frac{T^a T^b}{s-m_t^2} 		&\hspace{0.5cm} &	\mathbb{N}^{\mu\nu}_s = \gamma^\mu (\slashed p_1 + \slashed p_2+m_t)\gamma^\nu [g_V + g_A \gamma_5] \\
C_t = -\frac{T^b T^a}{t-m_t^2} 		& &	\mathbb{N}^{\mu\nu}_t = \gamma^\nu (\slashed p_1 - \slashed p_3+m_t) \gamma^\mu [g_V + g_A \gamma_5] \\
C_u = \frac{i T^c f^{abc} }{u-M^2} 	& & \mathbb{N}^{\mu\nu}_u = \left[ \eta^{\mu\nu} (\slashed p_2 + \slashed p_3 ) - 2 p_2^\mu \gamma^\nu - 2 p_3^\nu \gamma^\mu \right] (g_V + g_A \gamma_5).
\end{eqnarray}
In evaluating the cross section, we drop the small  corrections of order $m_t / M$ or $m_t / s$.  The collinear behavior manifests
as large $\log$s of the form $\log s / m_t^2$ or $\log (1 - M^2/s + m_t^2/s)$.

\paragraph{Counter-term}

As in the massless case, the collinear $\log$s have been absorbed into the top PDF and are subtracted from the hard matrix cross section by the
counter-term: 
\begin{equation}
\sigma_{gt}^{CT}(s) = -\frac{\alpha_s (g_V^2 + g_A^2) N_c C_F^2}{2 N_c (N_c^2 - 1) M^2} 
~\log\frac{\mu^2}{m_t^2}~ \left( 2 \tau^3 - 2\tau^2 + \tau \right).
\end{equation}
Combining the counter-term with the cross section for $t g \rightarrow t G$, we find that the collinear $\log$s all cancel, leaving behind
a finite contribution,
\begin{eqnarray}
\sigma^\text{finite}_{\text{real t}} &=& \frac{\alpha_s}{4\pi} \frac{\sigma_0}{2 M^2 C_F} \Bigg( 
C_F \bigg\{ -1 + 4\tau -3\tau^3 + \left[ -2\tau + 4\tau^2 - 4\tau^3 \right] \left( \log\frac{s\mu^2}{(s-M^2 + m_t^2)^2} \right) \bigg\} \nonumber\\&&
+ N_c  \left\{ 4\tau^3 - 7\tau^2 + 8\tau - 5 + (4 \tau^2 + 4\tau) \log \tau  + (4\tau^2 - 4\tau^3) \log\left(\frac{s-M^2 + m_t^2}{s-M^2- m_t^2} \right)  \right\} 
\Bigg). \nonumber\\
\label{eq:massive:final}
\end{eqnarray}
Once again the expression for $\sigma_{\text{real~$\bar t$}}^{\text{finite}}$ is identical.

\paragraph{Comparison to Light Quarks}

In the massless case, $\mathcal O(\epsilon)$ terms which multiply $1/\epsilon$ poles contribute to the finite cross section, 
but are absent in the top quark contribution.
As a result we observe that although the coefficients of the logarithms match the expression in~(\ref{eq:massless:final}), 
the simple polynomials in $\tau$ do not.
While these artifacts of the m-ACOT scheme markedly change the quark emission cross section, the effect on the inclusive cross section is small: 
the real quark contributions are themselves a small correction to the leading-logarithm cross section.


\section{NLO KK Gluon Cross Section}
\label{sec:results}

We assemble the real and virtual corrections into the full NLO+NLL cross section at ${\mathcal O} (\alpha_s)$ and 
${\mathcal O} (g_L^2 + g_R^2)$ and examine the theoretical predictions for KK gluon production as a function of its mass
at a 100 TeV proton-proton collider.  The rate for $pp \rightarrow G + X$ at NLO is computed by convolving the hard scattering
matrix elements with the appropriate PDFs,
\begin{eqnarray}
\sigma ( pp \rightarrow G + X ) &=& \sum_q
 \int_0^1\! dx_1 dx_2 \bigg( f_q(x_2,Q^2) f_{\bar q}(x_1,Q^2) \sigma_{q \bar q}
 + f_{ q}(x_1,Q^2) f_g(x_2,Q^2)  \sigma_{q g}  \nonumber \\ & & ~~~~~~~~~~~~~~~~~~~
 + f_{ \bar q}(x_1,Q^2) f_g(x_2,Q^2)  \sigma_{\bar q g} 
 + \big\{ x_1 \leftrightarrow x_2\big\} \bigg).	
 \label{eq:hadroniccrosssection}
\end{eqnarray}
The integral over the momentum fractions $x_1$ and $x_2$ is more conveniently transformed into one over $\tau$ and $y$:
\begin{equation}
x_1 \equiv \sqrt{\frac{\tau_0}{\tau}} e^y,  ~~~~x_2 \equiv \sqrt{\frac{\tau_0}{\tau}} e^{-y},~~~~\text{where}~~~~
\tau_0 \equiv \frac{M^2}{S},  ~~~~ \tau = \frac{M^2}{x_1 x_2 S} = \frac{\tau_0}{x_1 x_2}, 
\end{equation}
where $S$ is the $pp$ center of mass energy and with 
\begin{equation}
\int_0^1 dx_1 dx_2 \rightarrow \int_{1}^{\tau_0}\! d\tau \int_{-y_0}^{y_0}\! dy  \frac{\tau_0}{\tau^2}.
\label{eq:intty}
\end{equation}
The limits on the $y$ integral are:
\begin{equation}
y_0(\tau) = \frac{1}{2} \log \frac{\tau_0}{\tau}.
\end{equation}

The hard scattering cross section $\sigma_{\bar q q}$ is given by:
\begin{equation}
\sigma_{\bar q q} = \sigma_\text{LO} + \sigma_{\text{real}}^{\text{finite}} + \sigma_{\text{virt}}^{\text{finite}}.
\end{equation}
which are found in Equations~(\ref{eq:sigmaLO}),~(\ref{eq:virtual:finite}), and~(\ref{eq:real:finite}), respectively.
For the five light flavors of quark ($q=u,d,s,c,b$), the expressions for $\sigma_{q g}$ and $\sigma_{\bar q g}$
are given by Eq.~(\ref{eq:massless:final}).

The m-ACOT scheme dictates that we retain the top quark mass in the process $t g \rightarrow t G$, which changes the limits of integration on $\tau$
of Equation~(\ref{eq:intty}) such that its maximum occurs when the final state top and KK gluon are produced with no additional momentum, shifting it
away from 1 to:
\begin{eqnarray}
\tau_\text{max} &=& \frac{M^2}{s_\text{min}} = \frac{M^2}{(M+m_t)^2}.
\end{eqnarray}
The cross sections $\sigma_{t g}$ and $\sigma_{\bar t g}$ are both given by Equation~(\ref{eq:massive:final}).

\subsection{Sample RS Models}

We consider four illustrative sets of couplings based on popular RS models:
\begin{enumerate}
\item {\bf ``Anarchic"}: 
Based on an RS model with flavor-anarchic Higgs Yukawa couplings, which suggests particular bulk masses
for the various quarks, predicting that their couplings $g_L$ and $g_R$ 
(for quarks $\{ u, d, s, c, b, t \}$) are:
\begin{align}
\begin{array}{c} g_L \\  g_R \end{array} 
\; = \;
g_s \times \bigg\{ \begin{array}{l}  ( -0.2, ~-0.2, ~-0.2, ~-0.2, ~+1.0 ,~+1.0 ) \\ ( -0.2, ~-0.2, ~-0.2, ~-0.2, ~-0.2, ~+4.0 ) \end{array}.
\end{align}
\item {\bf ``Positive Anarchic":}
Small changes to the quark couplings can change the relative importance of the NLO corrections. 
We demonstrate this by modifying the $g_{L,R}^{(q)}$ couplings of the anarchic model such that they are positive:
\begin{align}
\begin{array}{c} g_L \\  g_R \end{array} 
\; = \;
g_s \times \bigg\{ \begin{array}{l}  ( 0.2, ~~0.2, ~~0.2, ~~0.2, ~~1.0 , ~~1.0 ) \\  (0.2, ~~0.2, ~~0.2, ~~0.2, ~~0.2, ~~4.0 ) \end{array}.
\end{align}
\item {\bf ``Symmetric":}
If one ignores the possibility for a geographic realization of quark flavor, one can engineer equal vector-like couplings for all quarks:
\begin{align}
g_L \; = \; g_R \; = \; 0.5 \times g_s.
\end{align}
\item {\bf ``Top-philic"}: If the only composite state is the right-handed top quark \cite{Lillie:2007hd}, there is typically a $\rho$-like state with 
large coupling to it, and very suppressed couplings to the light quarks:
\begin{align}
g^t_R \; \simeq 2\pi; ~~\text{all other couplings $\simeq$ zero.}
\end{align}
\end{enumerate}

\subsection{NLO Production Cross Section}

\begin{figure}[t]
\centering
\includegraphics[scale=1.0]{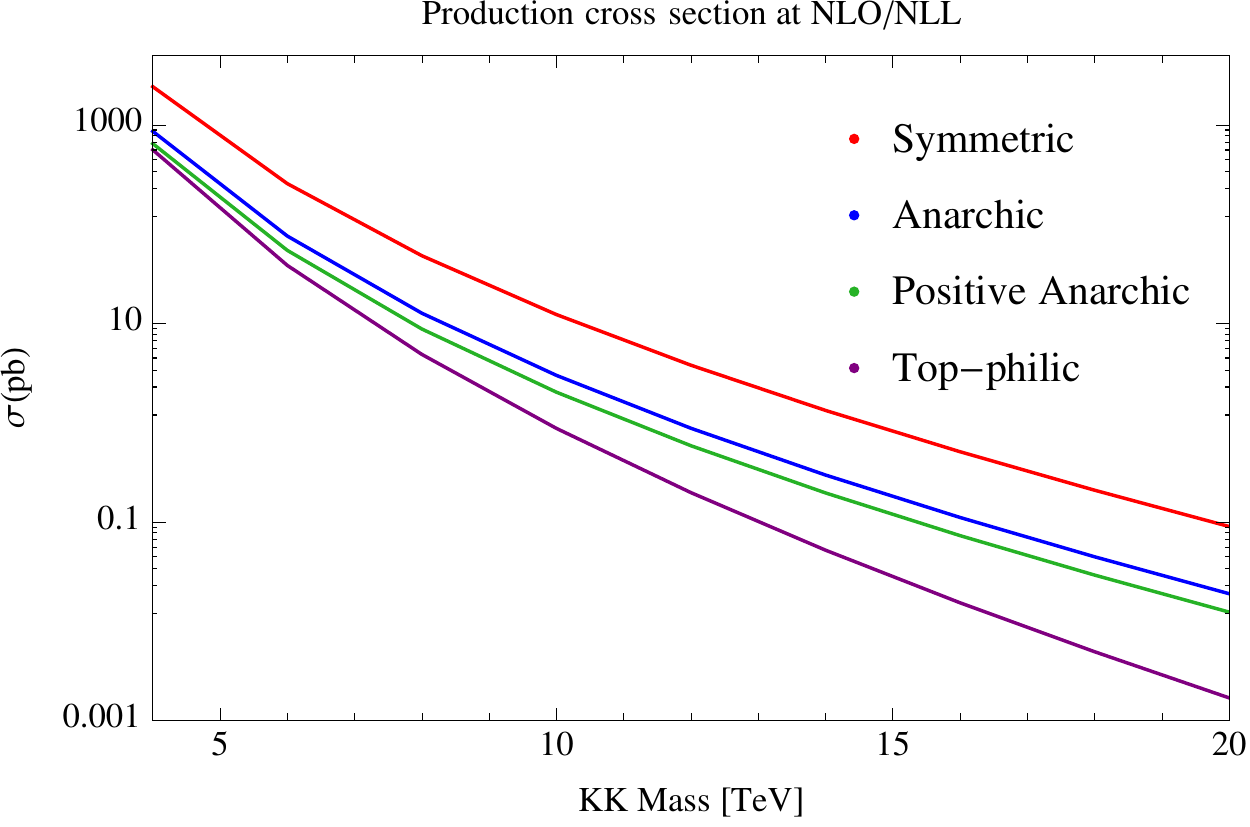}
\caption{Inclusive production cross section at NLO/NLL as a function of KK gluon mass $M$, with the coupling constants from the example
models discussed in the text.}
\label{figure:plots1}
\end{figure}

\begin{figure}[t]
\centering
\includegraphics[scale=.970]{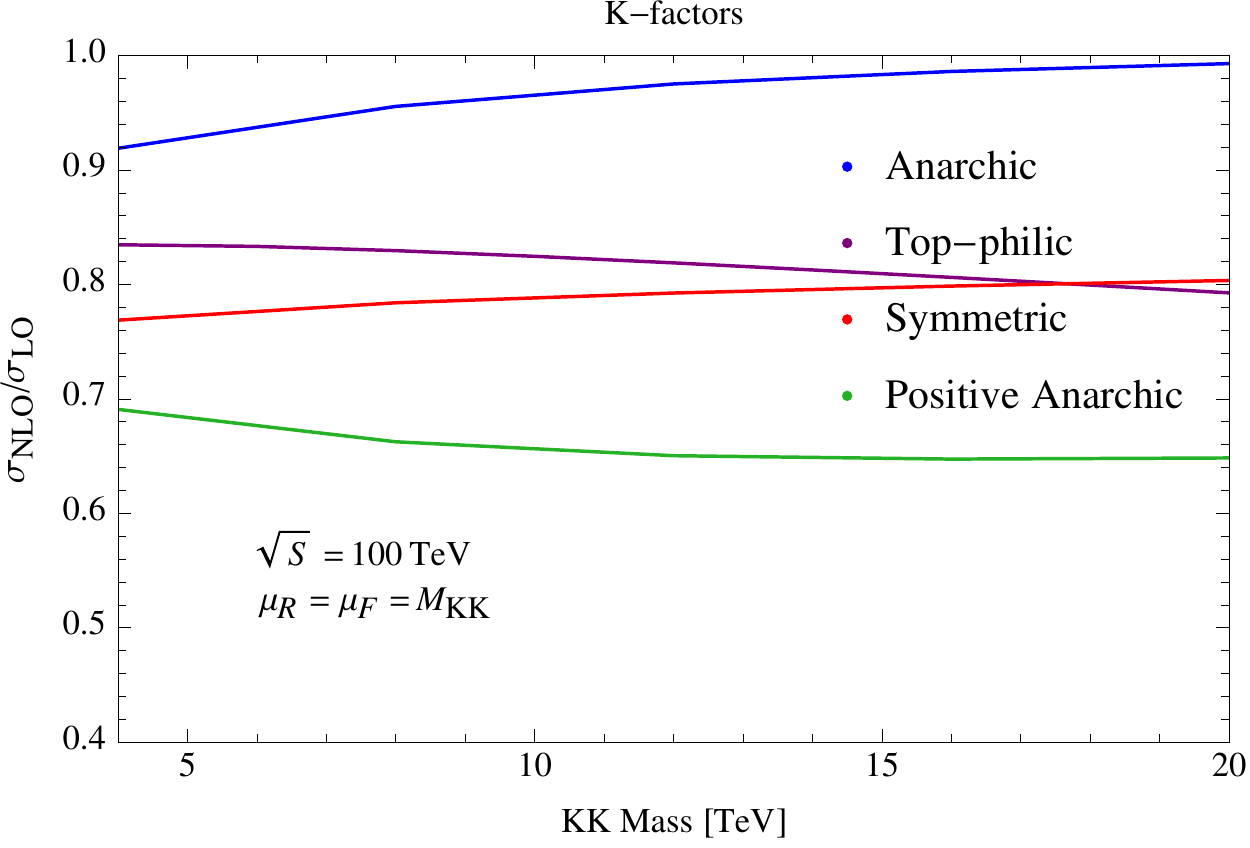}
\caption{$K$-factors as a function of mass, for the representative models discussed in the text. The $\sigma_\text{NLO}$ shown above includes the NLL and NLO contributions to the cross section.}
\label{figure:plots1b}
\end{figure}

For our four benchmark coupling sets, we evaluate the NLO production cross section at a $\sqrt{S} = 100$~TeV $pp$ collider.  We perform integrals
numerically using the
VEGAS package, together with the 6-flavor PDFs generated by NNPDF~3.0~\cite{Ball:2014uwa}. 
At $\sqrt{S} = 100$ TeV and $x\sim 0.1$, the uncertainty of the PDF luminosities can exceed $10\%$~\cite{Rojo:2016kwu}.

In Figure~\ref{figure:plots1}, we show results for the cross section with the NLO and NLL corrections included. 
Cross sections fall from around $\sim 1$~nb for masses around 5 TeV to around $\sim 100$~fb for masses around 20 TeV, and depend strongly on the
model determining the couplings. Despite the potentially strong coupling to the top quark, the small top quark PDF at $\sqrt{S} = 100$~TeV causes the light quarks to dominate the cross section in many models.
To demonstrate the relative importance of the NLO corrections, in Figure~\ref{figure:plots1b}
we plot for each coupling choice the $K$-factor, defined as $\sigma_\text{NLO} / \sigma_\text{LO}$. For both plots we have set the factorization scale $\mu_F$, and
the renormalization scale $\mu_R$, to the KK gluon mass $M$.  We observe that the $K$-factor is typically below one, and is as low as $\sim 0.7$ for the positive anarchic model.

\begin{figure}[h]
\centering
\includegraphics[scale=1.0]{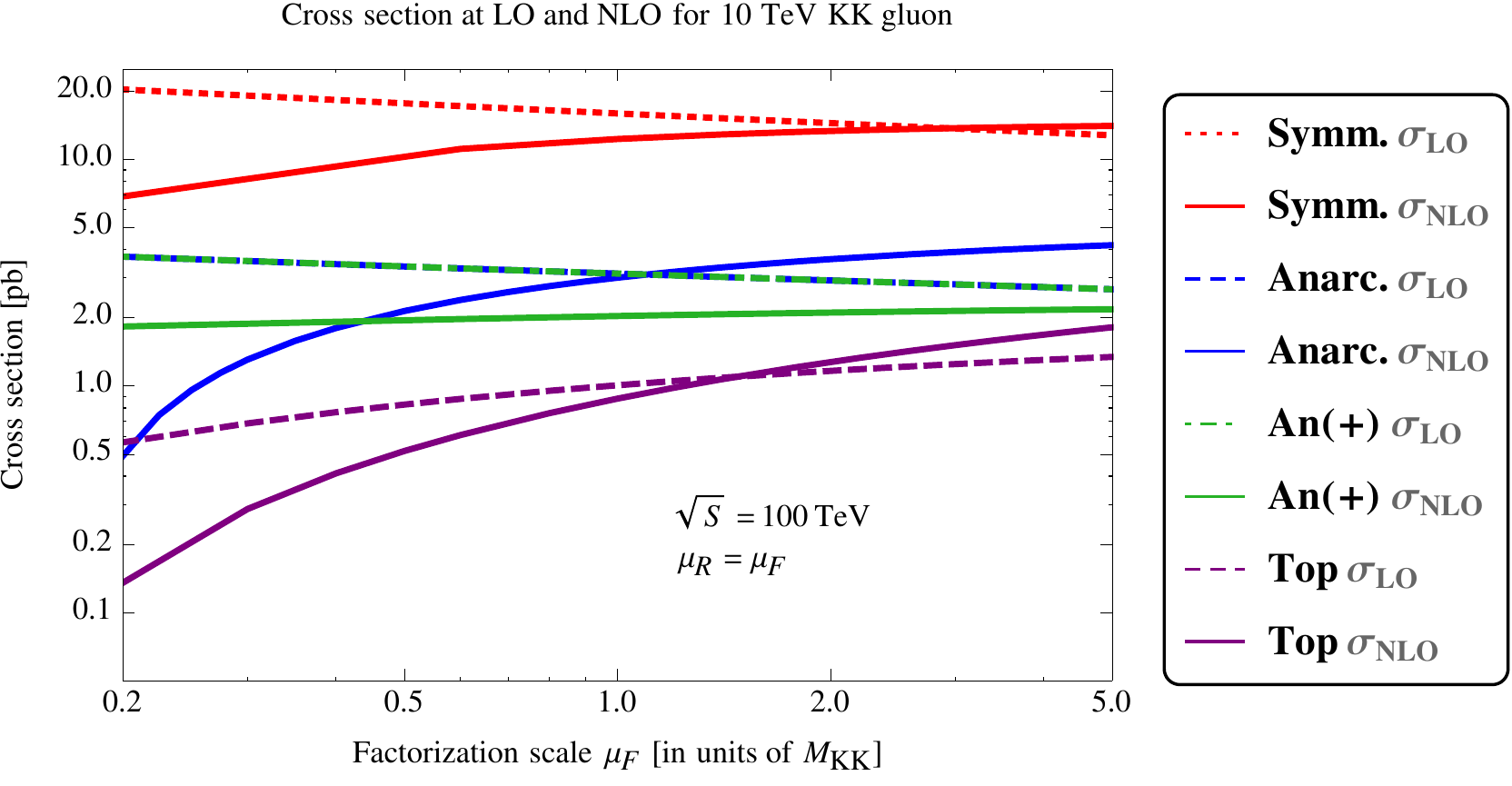}
\caption{Scale-dependence of the LO and NLO/NLL cross section for 10 \TeV\ KK gluon production.}
\label{figure:plots2}
\end{figure}

In Figure~\ref{figure:plots2} we examine the dependence on the scale $\mu = \mu_F = \mu_R$ of the LO and NLO cross sections for 
production of a 10 \TeV\ KK gluon in our various coupling scenarios.  We observe that the NLO scale dependence is somewhat stronger than the
LO scale dependence, indicating that LO scale variation would likely underestimate the uncertainty from higher order contributions, and 
that it would be useful to
consider NNLO corrections in the future.  If one uses the ``rule of thumb" variation between $M/2$ to $2M$ to estimate the scale uncertainty, it yields
an estimate of around $-30\%$ to $+17\%$ for the anarchic model, and smaller uncertainties for our other benchmark coupling choices.
Also evident in the figure is the fact that the anarchic and positive-anarchic models (whose couplings differ only by a sign) have the same LO
cross sections, but are distinguished at NLO where interference results in sensitivity to the relative signs of the couplings.

\section{Conclusion}

A 100 \TeV\ proton collider with a few $\text{fb}^{-1}$ integrated luminosity is able to produce KK gluons with $\mathcal O(10~\TeV)$ masses.  Such particles occur
commonly in theories of strong dynamics, such as the duals to the 5D weakly-coupled RS models.  KK gluons are perhaps the most effective diagnostics of
such models, with relatively large coupling to quarks.  In many such constructions, the coupling to the top quark is particularly strong.  Understanding the
results of experimental searches requires precise theoretical predictions for their production rates.  In this work, we have computed higher order corrections
to their production in a six-flavor scheme which treats the top quark as a parton.

In agreement with earlier studies \cite{Chivukula:2013xla,Zhu:2012um}, we find that the NLO corrections are typically negative for the canonical scale choice $\mu_F = \mu_R = M$, leading to
$K$-factors of order $0.7$, depending on the pattern of coupling to the quarks.  The NLO/NLL calculations exhibit somewhat more scale dependence
(of order $20\%$) compared to the LO approximation, and are thus important to estimate this theoretical systematic uncertainty.

A detailed experimental study considering high mass resonances decaying to a pair of top quarks at 100 TeV \cite{Auerbach:2014xua} assumes
10~ab$^{-1}$ of integrated luminosity.  They find that $\sigma \times$~BR of about 4 fb can probed by such a machine, leading to the conclusion that
KK gluons of masses up to about 20 TeV can be discovered.  Our higher order corrections suggest a more accurate estimate would be more like
18.5~TeV.  While this is less, it nonetheless argues that such a machine offers an unparalleled opportunity to probe strongly coupled theories.

\section*{Acknowledgments}

TMPT is grateful for conversations with S. Chivukula and especially A. Farzinnia concerning their work in Ref.~\cite{Chivukula:2013xla}.
This research was supported in part by NSF grant PHY-1316792 and by the University of California, Irvine through a Chancellor's Fellowship.

\appendix

\section{Integrals}

\subsection{Passarino-Veltman Decomposition} \label{section:handloop}

We follow the Passarino-Veltman reduction as described in~\cite{Ellis:2011cr}. 
In the loop diagrams of Section~\ref{section:triangle}, we need only the three $C_0$ functions defined below:
\begin{eqnarray}
C_0[p_1^2,p_2^2,q^2;0,0,0] &=& \mu^{2\epsilon} \int\!\frac{d^d k}{(2\pi)^d} \frac{1}{k^2 (k+p)^2 (k+q)^2} \\
C_0[p_1^2,p_2^2,q^2;0,0,M^2] &=& \mu^{2\epsilon} \int\!\frac{d^d k}{(2\pi)^d} \frac{1}{k^2 (k+p)^2 \left[ (k+q)^2 - M^2 \right]}  \\
C_0[p_1^2,p_2^2,q^2;0,M^2,0] &=& \mu^{2\epsilon} \int\!\frac{d^d k}{(2\pi)^d} \frac{1}{k^2 \left[(k+p)^2 - M^2 \right] (k+q)^2} .
\end{eqnarray}
From now on we suppress the momenta inputs to $C_0[\ldots]$. Note that $p_1^2=p_2^2=0$, and $q^2=M^2$ for all diagrams in Section~\ref{section:triangle}.

The first function, $C_0[0,0,0]$, corresponds to ``Diagram~A," where only massless particles run in the loop. This function is given in Appendix~E of~\cite{Ellis:2011cr}:
\begin{equation}
C_0[0,0,0] = \frac{i}{(4\pi)^2} \frac{1}{M^2} \left( \left[\frac{1}{\bar\epsilon^2} -\frac{\pi^2}{12} \right] + \frac{1}{\bar\epsilon} \log\frac{-\mu^2}{M^2} + \frac{1}{2} \log^2\frac{-\mu^2}{M^2} \right).
\end{equation}
Many authors, including~\cite{Ellis:2011cr}, include this factor of $-\pi^2/12$ in their definition of $1/\epsilon^2$. This choice has no effect on how the total cross section is written, because all factors of $1/\epsilon^2$ cancel each other.

\paragraph{Derivation of $C_0[0,0,M^2]$}
Diagrams B and C both use $C_0[0,0,q^2;0,0,M^2]$.
\begin{eqnarray}
C_0[0,0,M^2] &=& \mu^{2\epsilon} \int\!\frac{d^d k}{(2\pi)^d} \frac{1}{k^2 (k+p)^2 \left[ (k+q)^2 - M^2 \right]} \\
C_0[0,0,M^2] &=& \mu^{2\epsilon} \int_0^1\! dx dy dz \delta(1-x-y-z) \int\! \frac{d^dk}{(2\pi)^d} \frac{2!}{\mathcal D^3} ; \\
\mathcal D &\equiv& x k^2 + y (k+p)^2 + z (k+q)^2 - zM^2 \\
&=& k^2 + 2k \cdot (y p + z q) + yp^2 + z (q^2 - M^2) \nonumber\\
&\equiv& \ell^2 - (yp + qz)^2.
\end{eqnarray}
We define $\ell = k + yp + zq$ and use the on-shell conditions to simplify $p^2$ and $q^2$.  We define $\Delta$ such that:
\begin{eqnarray}
\Delta &\equiv& M^2 (yz + z^2). \\
C_0[0,0,M^2] &=& \mu^{2\epsilon} \int_0^1\! dz \int_0^{1-z} \! dy \int\! \frac{d^d\ell}{(2\pi)^d} \frac{2}{(\ell^2 - \Delta )^3} \\
&=& \mu^{2\epsilon} \frac{-i}{(4\pi)^2} \left( \frac{4\pi}{M^2} \right)^\epsilon \Gamma(1+\epsilon) \int_0^1 \! dz \int_0^{1-z} \! dy \left[ \frac{1}{yz + z^2} \right]^{1+\epsilon} \frac{1}{M^2} \\
&=& \frac{-i}{M^2 16\pi^2} \left( \frac{4\pi \mu^2}{M^2} \right)^\epsilon \Gamma(1+\epsilon) \int_0^1\!dz \left(\frac{1}{z} \right)^{1+\epsilon} \left[ \frac{1^{-\epsilon} - z^{-\epsilon} }{-\epsilon} \right] \\
&=& \frac{i}{16\pi^2} \frac{\Gamma(1+\epsilon)}{M^2 \epsilon} \left( \frac{4\pi \mu^2}{M^2} \right)^\epsilon \left[ \frac{\Gamma(-\epsilon) \Gamma(1) }{\Gamma(1-\epsilon)} - \frac{ \Gamma(-2\epsilon) \Gamma(1) }{\Gamma(1-2\epsilon) } \right] \\
C_0[0,0,M^2] &=& \frac{i}{(4\pi)^2} \left(\frac{-1}{2M^2} \right) \left[ \frac{1}{\bar\epsilon^2} + \frac{\log(\mu^2/M^2)}{\bar\epsilon} + \frac{1}{2} \log^2\frac{\mu^2}{M^2} + \frac{\pi^2}{12} \right].
\end{eqnarray}

\paragraph{Derivation of $C_0[0,M^2,0]$}
The loop integral of Diagram~D is distinct from the others, and is not IR divergent.
\begin{eqnarray}
C_0[0,M^2,0] &=& \mu^{2\epsilon} \int\!\frac{d^d k}{(2\pi)^d} \frac{1}{k^2 \left[ (k+p)^2 - M^2 \right] (k+q)^2 } \\
C_0[0,M^2,0] &=& \mu^{2\epsilon} \int_0^1\! dx dy dz \delta(1-x-y-z) \int\! \frac{d^dk}{(2\pi)^d} \frac{2!}{\mathcal D^3} ; \\
\mathcal D &\equiv& x k^2 + y (k+p)^2 - yM^2 + z (k+q)^2  \\
&=& k^2 + 2k \cdot(yp + zq) + y (p^2 - M^2) + z q^2 \\
&=& \ell^2 - \left[(yp + zq)^2 + y M^2 - z q^2 \right] \; =\; \ell^2 - \Delta,\\
\Delta &=&  M^2 (yz  + y - z + z^2) + i \varepsilon,
\end{eqnarray}
with $\ell^\mu = k^\mu + yp^\mu + zq^\mu$. The $i\varepsilon$ term is useful for keeping track of branch cuts in the polylogarithms that appear in the integral. This integral not UV divergent, and the IR divergences cancel. While $\epsilon$ may be used to regulate the IR divergences, this is not necessary. 
\begin{eqnarray}
C_0[0,M^2,0] &=&  \frac{-i}{(4\pi)^2} \left( \frac{4\pi \mu^2}{M^2} \right)^\epsilon \Gamma(1+\epsilon) \int_0^1 \! dz \int_0^{1-z} \! dy \frac{1}{M^2} \left[ \frac{1}{y(1+z) - z +z^2 - i\varepsilon} \right]^{1+\epsilon}  \\
&=& \frac{-i}{(4\pi)^2} \left( \frac{4\pi \mu^2}{M^2} \right)^\epsilon \frac{ \Gamma(1+\epsilon) }{M^2} \cdot \mathcal I  \\
\mathcal I &=& \int_0^1 \! dz \int_{-z(1-z)}^{1-z} \! \frac{dy'}{1+z} \left( \frac{1}{y' - i\varepsilon } \right)^{1+\epsilon}
 \; = \; \int_0^1 \! \frac{dz}{1+z} \mathcal I_y(z), \\
\mathcal I_y &=& \int_{+z(1-z)}^{1-z} \! dy' \left( \frac{1}{y' - i\varepsilon } \right)^{1+\epsilon} + \int_{-z(1-z)}^{z(1-z)} \! dy' \left( \frac{1}{y' - i\varepsilon } \right)^{1+\epsilon} \\
&=& \log \left(\frac{1-z}{z(1-z)} \right) + i \pi + \mathcal O(\epsilon) + \mathcal O(\varepsilon) . \\
\mathcal I &=& \int_0^1 \frac{dz}{1+z} \left[ \log \left(\frac{1}{z} \right) + i \pi \right] 
\; = \; - \int_0^1 \! dz \frac{[\log(-z)]^\star }{1+z} \; = \; - \int_1^2 \! dz' \frac{[\log(1-z')]^\star }{z'} \nonumber\\
\end{eqnarray}
Integrating this last term requires the use of polylogarithms, $\text{Li}_n(x)$, and their recursive relationship:
\begin{align}
\frac{d}{dx} \text{Li}_n(x) \;=\; \frac{1}{x} \text{Li}_{n-1} (x) && \text{Li}_1(x) \;=\; - \log(1-x).
\end{align}
Replacing $\text{Li}_2 (1)$ with its analytic expression produces the following expression for $C_0[0,M^2,0]$.
\begin{eqnarray}
C_0[0,M^2,0] &=&\frac{i}{(4\pi)^2} \frac{1}{M^2} \left[ \frac{\pi^2}{6} - \text{Li}_2 (2)  \right]^\star.
\end{eqnarray}
The real parts of the three scalar functions calculated above match the results from \emph{Package~X}~1.0.4~\cite{Patel:2015tea}.

\subsection{Scaleless Loop Integral} \label{section:scalelessloop}
One may show explicitly that the scaleless loop integral is zero by adding and subtracting a term with a nonzero mass. This splits the expression into a term that is strictly IR-divergent and a term that is strictly UV-divergent. 
\begin{eqnarray}
\int\!\frac{d^d \ell}{(2\pi)^2} \frac{1}{(l^2)^2} &=& \int\!\frac{d^d \ell}{(2\pi)^2} \Big[ \frac{1}{(l^2)^2} - \frac{1}{(\ell^2-M^2)^2} + \frac{1}{(\ell^2- M^2)^2} \Big]; \\
\Big[ \frac{1}{(\ell^2)^2} - \frac{1}{(\ell^2-M^2)^2} \Big] &=& \frac{\ell^4-2 \ell^2 M^2 + M^4 - \ell^4}{\ell^2(\ell^2-M^2)} \\
&=&\int_0^1\!\text{d}y \frac{y(1-y) \Gamma(4) (M^4 - 2\ell^2 M^2)}{(\ell^2 - y M^2)^4}.
\end{eqnarray}
The Euler beta function simplifies the integral in $y$. 
\begin{eqnarray}
\lefteqn{\int\!\frac{d^d \ell}{(2\pi)^2} \int_0^1\!\text{d}y \frac{y(1-y) \Gamma(4)(M^4 - 2\ell^2 M^2)}{(\ell^2 - y M^2)^4}} \nonumber\\ &=& \frac{\Gamma(4) i}{(4\pi)^{d/2} \Gamma(4) } \int_0^1\!\text{d}y (1-y)y 
\Big[\frac{M^4 \Gamma(2+\epsilon)}{(y M^2)^{2+\epsilon} } + \frac{2 M^2 \Gamma(1+\epsilon)(2-\epsilon)}{(yM^2)^{1+\epsilon} } \Big] \\
&=& \frac{i \Gamma(1+\epsilon)}{(4\pi^2)} \Big(\frac{4\pi}{M^2}\Big)^\epsilon \Big[ \frac{(1+\epsilon) \Gamma(-\epsilon) \Gamma(2)}{\Gamma(2-\epsilon)} + \frac{(4-2\epsilon) \Gamma(1-\epsilon) \Gamma(2) }{ \Gamma(3-\epsilon) } \Big];
\end{eqnarray}
The scaleless integral is identically zero in the $\epsilon\rightarrow0$ limit:
\begin{eqnarray}
\int\!\frac{d^d \ell}{(2\pi)^2} \frac{1}{(\ell^2)^2} &=& \frac{i}{(4\pi)^2} \Big(\frac{4\pi}{M^2}\Big)^\epsilon \Big[ \frac{\Gamma(1+\epsilon) (1+\epsilon) \Gamma(-\epsilon)}{\Gamma(2-\epsilon) } + \frac{(4-2\epsilon) \Gamma(1+\epsilon) \Gamma(1-\epsilon) }{ \Gamma(3-\epsilon)} + \Gamma(\epsilon) \Big] \nonumber\\
&=& \frac{i}{(4\pi)^2} \Big(\frac{4\pi}{M^2}\Big)^\epsilon \Big[ \Big( -\frac{1}{\epsilon} + \gamma_E - 2 \Big) +  (2) + \Big( \frac{1}{\epsilon} -\gamma_E \Big) \Big] \\
\int\!\frac{d^d \ell}{(2\pi)^2} \frac{1}{(\ell^2)^2} &=& 0. 	\label{eq:scaleless}
\end{eqnarray}

\subsection{Tables of Phase Space Integrals}		\label{section:integraltables}
We use the tabulated integrals of~\cite{Beenakker:1988bq} to integrate over the angular coordinates. In the first case the KK gluon is produced with a massless gluon or quark; in the second case, with a top quark of mass $m_t$.

\begin{table}[h]
\begin{align*}
\int\! d\Theta\ 1 &= 2\pi 			& & \int\! d\Theta\ \frac{1}{t} = \frac{2\pi}{(s-M^2)} \frac{1}{\epsilon} \\
\int\! d\Theta\ t &= \pi(M^2-s) 		& & \int\! d\Theta\ \frac{1}{u} = \frac{2\pi}{(s-M^2)} \frac{1}{\epsilon} \\
\int\! d\Theta\ u &= \pi(M^2-s) 		& & \int\! d\Theta\ (u-M^2) = -\pi(M^2+s) \\
\int\! d\Theta\ t^2 &= \frac{2\pi}{3}(s-M^2)^2 	& & \int\! d\Theta\ \frac{1}{(u-M^2)^2} = \frac{2\pi}{s\cdot M^2} \\
\int\! d\Theta\ u^2 &= \frac{2\pi}{3}(s-M^2)^2 	& &  \int\! d\Theta\ \frac{1}{(u-M^2)} = \frac{2\pi}{(s-M^2)}\log(M^2/s)\\
\int\! d\Theta\ \frac{1}{tu} &= \frac{-4\pi}{(s-M^2)^2} \frac{1}{\epsilon} & &
\int\! d\Theta\ \frac{1}{t(u-M^2)} = \frac{-2\pi}{s(s-M^2)}\Big(\frac{1}{\epsilon}+ \log(M^2/s) \Big) 
\end{align*}
\caption{Table of integrals for massless quark or gluon emission.}
\label{figure:integrals:massless}
\end{table}

In the massive quark case, the following constants appear in the loop integrals shown in Table~\ref{figure:integrals:massive}:
\begin{eqnarray}
a &=& - \frac{(s-m_t^2)(s+m_t^2 - M^2)}{2 s} \\
b &=&  \frac{s-m_t^2}{2s} \sqrt{\left(s-m_t^2 -M^2 \right)^2 - 4 m_t^2 M^2} = -\beta_4 a\\
A &=& - \frac{(s-m_t^2)(s-m_t^2 + M^2)}{2 s} \\
B &=& - \frac{s-m_t^2}{2s} \sqrt{\left(s-m_t^2 -M^2 \right)^2 - 4 m_t^2 M^2} = - b \\
\sqrt{X} &=& \frac{\left(s-m_t^2\right)^2}{2 s} \sqrt{ (s-M^2 - m_t^2)^2 - 4M^2 m_t^2 } = (s-m_t^2) b.
\end{eqnarray}

\begin{table}
\begin{align*}
&\int\! d\Theta\ 1 = 2\pi 		&& \int\! d\Theta\ \frac{1}{t-m_t^2}\frac{1}{u-M^2} = \frac{\pi}{\sqrt{X}} \log\left[\frac{ aA-bB+\sqrt{X}}{aA-bB-\sqrt{X}}\right] \\
&\int\! d\Theta\ (t-m_t^2) = 2\pi a 	&& \int\! d\Theta\ (u-M^2) = 2\pi A  \\
&\int\! d\Theta\ \frac{1}{t-m_t^2} = \frac{\pi}{b} \log\frac{ a+b}{a-b}	&&\int\! d\Theta\ \frac{1}{u-M^2} = \frac{\pi}{B} \log\frac{ A+B}{A-B} \\
&\int\! d\Theta\ \left(\frac{1}{t-m_t^2}\right)^2 = \frac{2\pi}{a^2-b^2} 	&& \int\! d\Theta\ \left(\frac{1}{u-M^2}\right)^2 = \frac{2\pi}{A^2-B^2} 
\end{align*}
\caption{Table of angular integrals for massive quark emission.}
\label{figure:integrals:massive}
\end{table}

\bibliographystyle{utphys} 
\bibliography{KKGluon}

\end{document}